\documentstyle[12pt]{article}
\catcode`\@=11

\@addtoreset{equation}{section}
\def\theequation{\arabic{equation}}
\def\theequation{\thesection\arabic{equation}}

\def\NPB#1#2#3{{\it Nucl.~Phys.} {\bf{B#1}} (19#2) #3}
\def\PLB#1#2#3{{\it Phys.~Lett.} {\bf{B#1}} (19#2) #3}
\def\PRD#1#2#3{{\it Phys.~Rev.} {\bf{D#1}} (19#2) #3}

\def\@normalsize{\@setsize\normalsize{15pt}\xiipt\@xiipt
\abovedisplayskip 14pt plus3pt minus3pt%
\belowdisplayskip \abovedisplayskip
\abovedisplayshortskip  \z@ plus3pt%
\belowdisplayshortskip  7pt plus3.5pt minus0pt}
\def\small{\@setsize\small{13.6pt}\xipt\@xipt
\abovedisplayskip 13pt plus3pt minus3pt%
\belowdisplayskip \abovedisplayskip
\abovedisplayshortskip  \z@ plus3pt%
\belowdisplayshortskip  7pt plus3.5pt minus0pt
\def\@listi{\parsep 4.5pt plus 2pt minus 1pt
            \itemsep \parsep
            \topsep 9pt plus 3pt minus 3pt}}

\def\underline#1{\relax\ifmmode\@@underline#1\else
        $\@@underline{\hbox{#1}}$\relax\fi}
\@twosidetrue
\relax

\catcode`@=12

\catcode`\@=11

\def\section{\@startsection{section}{1}{\z@}{3.5ex plus 1ex minus
   .2ex}{2.3ex plus .2ex}{\large\bf}}
\def\thesection{\arabic{section}.}

\def\ps@headings{\def\@oddfoot{}\def\@evenfoot{}
\def\@oddhead{\hbox{}\hfill
        \makebox[.5\textwidth]{\raggedright\ignorespaces --\thepage{}--
        \hfill }}
\def\@evenhead{\@oddhead}
\def\subsectionmark##1{\markboth{##1}{}} }

\ps@headings

\catcode`\@=12

\relax

\def\figcap{\section*{Figure Captions\markboth
        {FIGURECAPTIONS}{FIGURECAPTIONS}}\list
        {Fig. \arabic{enumi}:\hfill}{\settowidth\labelwidth{Fig. 999:}
        \leftmargin\labelwidth
        \advance\leftmargin\labelsep\usecounter{enumi}}}
 \relax
\def\tablecap{\section*{Table Captions\markboth
        {TABLECAPTIONS}{TABLECAPTIONS}}\list
        {Table \arabic{enumi}:\hfill}{\settowidth\labelwidth{Table 999:}
        \leftmargin\labelwidth
        \advance\leftmargin\labelsep\usecounter{enumi}}}
 \relax
\def\reflist{\section*{References\markboth
        {REFLIST}{REFLIST}}\list
        {[\arabic{enumi}]\hfill}{\settowidth\labelwidth{[999]}
        \leftmargin\labelwidth
        \advance\leftmargin\labelsep\usecounter{enumi}}}
 \relax

\catcode`\@=11

\def\marginnote#1{}
\newcount\hour
\newcount\minute
\newtoks\amorpm
\hour=\time\divide\hour by60
\minute=\time{\multiply\hour by60 \global\advance\minute by-
\hour}
\edef\standardtime{{\ifnum\hour<12 \global\amorpm={am}%
    \else\global\amorpm={pm}\advance\hour by-12 \fi
    \ifnum\hour=0 \hour=12 \fi
    \number\hour:\ifnum\minute<100\fi\number\minute\the\amorpm}}
\edef\militarytime{\number\hour:\ifnum\minute<100\fi\number\minute}
\def\draftlabel#1{{\@bsphack\if@filesw {\let\thepage\relax
  \xdef\@gtempa{\write\@auxout{\string
    \newlabel{#1}{{\@currentlabel}{\thepage}}}}}\@gtempa
    \if@nobreak \ifvmode\nobreak\fi\fi\fi\@esphack}
     \gdef\@eqnlabel{#1}}
\def\@eqnlabel{}
\def\@vacuum{}
\def\draftmarginnote#1{\marginpar{\raggedright\scriptsize\tt#1}}
\def\draft{\oddsidemargin -.5truein
        \def\@oddfoot{\sl preliminary draft \hfil
        \rm\thepage\hfil\sl\today\quad\militarytime}
        \let\@evenfoot\@oddfoot \overfullrule 3pt
        \let\label=\draftlabel
        \let\marginnote=\draftmarginnote
   
\def\@eqnnum{(\theequation)\rlap{\kern\marginparsep\tt\@eqnlabel}%
\global\let\@eqnlabel\@vacuum}  }
\def\preprint{\twocolumn\sloppy\flushbottom\parindent 1em
        \leftmargini 2em\leftmarginv .5em\leftmarginvi .5em
        \oddsidemargin -.5in    \evensidemargin -.5in
        \columnsep 15mm \footheight 0pt
        \textwidth 250mmin      \topmargin  -.4in
        \headheight 12pt \topskip .4in
        \textheight 175mm
        \footskip 0pt
        
\def\@oddhead{\thepage\hfil\addtocounter{page}{1}\thepage}
        \let\@evenhead\@oddhead \def\@oddfoot{} \def\@evenfoot{}  }
\def\titlepage{\@restonecolfalse\if@twocolumn\@restonecoltrue\onecolumn
     \else \newpage \fi \thispagestyle{empty}\c@page\z@
        \def\thefootnote{\fnsymbol{footnote}} }
\def\endtitlepage{\if@restonecol\twocolumn \else  \fi
        \def\thefootnote{\arabic{footnote}}
        \setcounter{footnote}{0}}  
\catcode`@=12
\relax

\def\ps@headings{\def\@oddfoot{}\def\@evenfoot{}
\def\@oddhead{\hbox{}\hfill
        \makebox[.5\textwidth]{\raggedright\ignorespaces --\thepage{}--
        \hfill }}
\def\@evenhead{\@oddhead}
\def\subsectionmark##1{\markboth{##1}{}} }

\ps@headings

\relax

\def\firstpage#1#2#3#4#5#6{
\begin{document}
\begin{titlepage}
\nopagebreak
\title{\begin{flushright}
        \vspace*{-1.8in}
        {\normalsize CERN-TH/98-212}\\[-9mm]
        {\normalsize CPTH-S616.0698}\\[-9mm]
        {\normalsize LPTHE-ORSAY 98/42}\\[-9mm]
        {\normalsize ROM2F-98/21}\\[-9mm]
        {\normalsize hep-th/9807011}\\[4mm]
\end{flushright}
\vfill {#3}}
\author{\large #4 \\[1.0cm] #5}
\maketitle
\vskip -7mm     
\nopagebreak 
\begin{abstract} {\noindent #6}
\end{abstract}
\vfill
\begin{flushleft}
\rule{16.1cm}{0.2mm}\\[-3mm]
$^{\star}${\small Research supported in part by the EEC under TMR contract 
ERBFMRX-CT96-0090.}\\[-3mm] 
$^{\dagger}${\small Laboratoire associ{\'e} au CNRS-URA-D0063.}\\ 
CERN-TH/98-212\\ July 1998
\end{flushleft}
\thispagestyle{empty}
\end{titlepage}}

\def\simlt{\stackrel{<}{{}_\sim}}
\def\simgt{\stackrel{>}{{}_\sim}}
\newcommand{\dal}{\raisebox{0.085cm} {\fbox{\rule{0cm}{0.07cm}\,}}}
\newcommand{\dt}{\partial_{\langle T\rangle}}
\newcommand{\dtbar}{\partial_{\langle\overline{T}\rangle}}
\newcommand{\al}{\alpha^{\prime}}
\newcommand{\mst}{M_{\scriptscriptstyle \!S}}
\newcommand{\mpl}{M_{\scriptscriptstyle \!P}}
\newcommand{\dv}{\int{\rm d}^4x\sqrt{g}}
\newcommand{\lv}{\left\langle}
\newcommand{\rv}{\right\rangle}
\newcommand{\ph}{\varphi}
\newcommand{\abar}{\overline{a}}
\newcommand{\sbar}{\,\overline{\! S}}
\newcommand{\xbar}{\,\overline{\! X}}
\newcommand{\fbar}{\,\overline{\! F}}
\newcommand{\zbar}{\overline{z}}
\newcommand{\dbar}{\,\overline{\!\partial}}
\newcommand{\tbar}{\overline{T}}
\newcommand{\taubar}{\overline{\tau}}
\newcommand{\ubar}{\overline{U}}
\newcommand{\ybar}{\overline{Y}}
\newcommand{\phb}{\overline{\varphi}}
\newcommand{\cm}{Commun.\ Math.\ Phys.~}
\newcommand{\prl}{Phys.\ Rev.\ Lett.~}
\newcommand{\pr}{Phys.\ Rev.\ D~}
\newcommand{\pl}{Phys.\ Lett.\ B~}
\newcommand{\ibar}{\overline{\imath}}
\newcommand{\jbar}{\overline{\jmath}}
\newcommand{\np}{Nucl.\ Phys.\ B~}
\newcommand{\F}{{\cal F}}
\renewcommand{\L}{{\cal L}}
\newcommand{\A}{{\cal A}}
\newcommand{\e}{{\rm e}}
\newcommand{\be}{\begin{equation}}
\newcommand{\ee}{\end{equation}}
\newcommand{\ba}{\begin{eqnarray}}
\newcommand{\ea}{\end{eqnarray}}
\newcommand{\dslash}{{\not\!\partial}}
\newcommand{\gsi}{\,\raisebox{-0.13cm}{$\stackrel{\textstyle >}{\textstyle\sim}$}\,}
\newcommand{\lsi}{\,\raisebox{-0.13cm}{$\stackrel{\textstyle <}{\textstyle\sim}$}\,}
\date{}
\firstpage{3118}{IC/95/34} {\large\bf Supersymmetry breaking, open strings and
M-theory}  {I. Antoniadis$^{\,a,b}$, E. Dudas$^{\,b,c}$ and A. Sagnotti$^{\,d}$}
{\normalsize\sl
$^a$ Centre de Physique Th{\'e}orique,  Ecole Polytechnique, {}F-91128 Palaiseau,
France\\[-3mm]
\normalsize\sl$^b$ TH-Division, CERN, CH-1211 Geneva 23, Switzerland\\[-3mm]
\normalsize\sl $^c$  LPTHE$^\dagger$, B{\^a}t. 211, Univ. Paris-Sud, F-91405 Orsay,
France\\[-3mm]
\normalsize\sl$^{d}$ Dipartimento di Fisica, Universita di Roma ``Tor Vergata'', INFN,
Sezione di Roma,\\[-3mm]\normalsize\sl Via della Ricerca Scientifica 1, 00133 Roma,
Italy} 
{We study supersymmetry breaking by Scherk-Schwarz compactifications in type I
string theory. While in the gravitational sector all mass splittings are  proportional
to a (large) compactification radius, supersymmetry remains  unbroken for the massless
excitations of D-branes orthogonal to the large dimension. In this sector,
supersymmetry breaking can
then be mediated by gravitational interactions alone, 
that are expected to be suppressed by powers of the
Planck mass. The mechanism is non perturbative from the heterotic viewpoint and requires
a compactification radius at intermediate energies  of order
$10^{12}-10^{14}$ GeV. This can also explain the value of Newton's constant if the
string scale is close to the unification scale, of order $10^{16}$ GeV.}
\section{Introduction}

In perturbative string theory supersymmetry breaking through compactification  relates
the breaking scale to the size of a compact dimension. In weakly coupled heterotic
strings, one thus obtains the (tree-level) relation:
\be m_{3/2}=m_{1/2}\sim R^{-1}\, ,
\label{sh}
\ee where $m_{3/2}$ and $m_{1/2}$ are the gravitino and gaugino masses, 
while $R$ is the radius of the extra dimension. Therefore, phenomenologically acceptable
soft masses ask for a very large radius, in the TeV$^{-1}$ region. However, since
Standard Model gauge bosons feel the extra dimension, only in special models can one
avoid large corrections to the gauge couplings and accommodate the existing
phenomenology \cite{a,AMQ}. 
The resulting scalar masses $m_0$ are then in general model dependent. In
the simplest case, chiral families live on boundaries of space orthogonal to the TeV
dimension, and thus do not feel it; they correspond to $N=1$ twisted states and satisfy
$m_0=0$. This guarantees flavor universality, avoids potential problems with proton
decay and makes model building easier and more natural.

The Scherk-Schwarz mechanism provides an elegant realization of supersymmetry breaking by
compactification in field theory \cite{SS}. In the simplest case of circle compactification, it
amounts to allowing the higher dimensional fields to be periodic around the circle up to
an R-symmetry transformation. The Kaluza-Klein momenta of the various fields 
are correspondingly shifted in
a way proportional to their R-charges. The R-transformations are
actually restricted to discrete subgroups, as are all global internal symmetries of
supergravity models, since the massive excitations lead to effective Dirac-like
quantizations of the corresponding parameters. As a result, the scale of supersymmetry
breaking is quantized in units of the compactification scale. Modular invariance
dictates the extension of this mechanism to the full perturbative spectrum in models of
oriented closed strings \cite{r,KP,a}.

Breaking supersymmetry by compactification rather than dynamically
via gaugino condensation \cite{DIN} has the obvious advantage of calculability. 
However, in the perturbative
heterotic string this breaking appears to be too restrictive, and it is interesting to
inquire whether other possibilities can be realized within different perturbative
string descriptions. 

In this work we study supersymmetry breaking by compactification in the type I theory of
open and closed strings. This is closely related to the type-IIB theory of oriented
closed strings \cite{S}, and can also describe the heterotic string
at strong coupling \cite{W1,PW,HW}.
In particular, we generalize the Scherk-Schwarz mechanism to open strings. Although the
gravitino  mass is still proportional to a compactification radius, we find that the
gaugino mass can actually decouple. This occurs whenever in the type I theory gauge
particles live on D-branes orthogonal to the large dimension. In the low-energy spectrum
supersymmetry breaking then affects only the bulk to lowest order, and thus
\be 
m_{3/2}\sim R^{-1}\ , \qquad\qquad m_{1/2}=0\quad ,
\label{sI}
\ee 
while on the brane it is mediated by gravitational interactions, expected to be
suppressed by powers
of the Planck mass $M_P$, so that effectively $m_{1/2}\simlt{\cal O}(m_{3/2}^2/M_P)$. As a
result, the compactification radius could be at an intermediate scale, $R^{-1} \sim
10^{12}-10^{14}$ GeV. These models do not require special conditions to avoid the 
large coupling problem, since  the large dimension is felt by matter fields only via
gravitational  interactions.

This scenario was suggested \cite{AQ} in the context of M-theory, 
where supersymmetry breaking
could be related to the radius of the eleventh dimension \cite{AQ,DG}. 
In fact,  as we will show in
Section 4, upon compactification to 9 dimensions there is an equivalent (dual)
description as a weakly coupled type I$^\prime$ theory. 
Models with an intermediate scale
compactification and with the breaking conditions (\ref{sI}) are certainly
non-perturbative when viewed from the heterotic side. They share many properties with
models of gaugino condensation, and may even provide a dual description of it if the
condensation scale is the M-theory scale and the generated superpotential is of order
one in Planck units.

This paper is organized as follows. In Section 2 we discuss gauge coupling unification
in type I theories and show that it is compatible with intermediate  scale
compactification consistently with the supersymmetry breaking masses
(\ref{sI}). In Section 3 we review the BPS spectrum in $N=1$ compactifications to nine
dimensions obtained either from M-theory on $S^1\times S^1/Z_2$ or from type $I$ (or
type $I^\prime$) string theory on $S^1$.  In Section 4 we discuss supersymmetry breaking
by the Scherk-Schwarz mechanism in the effective field theory and in closed
string theory. In Section 5 we generalize this mechanism to open strings and give
explicit examples in nine dimensions. We find two distinct possibilities. The first
extends to the type I string the ordinary Scherk-Schwarz compactification, and can also
be obtained by duality from the perturbative $SO(32)$ heterotic string in the limit
where the winding modes decouple from the gauge sector. 
It also defines type I theory at finite temperature, when the compactified coordinate
is identified with the (euclidean) time.
The second corresponds to a similar
compactification on the type I$^\prime$ side,  but now the
$SO(16)\times SO(16)$ gauge group lives on D8-branes orthogonal to the dimension
used to break supersymmetry. As a result, supersymmetry remains unbroken in the massless
gauge sector. By duality, this mechanism is equivalent to Scherk-Schwarz supersymmetry
breaking along the 11th dimension on the  M-theory side, and is non-perturbative from
the heterotic point of view. In Section 6 we compute the one-loop cosmological constant
and the scalar masses for the two models. 
In the second case, the corrections to both quantities from
open-string loops are exponentially suppressed in the large radius limit. This result
agrees with the field theory expectation that, in the absence of quadratic divergences,
supersymmetry breaking mediated by gravitational  interactions should be suppressed by
powers of the Planck mass. In Section 7 we generalize the construction to six
dimensions and present an explicit chiral model with broken $(1,0)$ supersymmetry.

\section{Unification and supersymmetry breaking in open strings and in M-theory}

In this Section, we consider unification constraints in type I and type I$^\prime$
models and corresponding scenarios of intermediate scale supersymmetry  breaking via 
compactification along a certain direction in the compact space.  In nine-dimensional
models, this direction can be identified with the eleventh dimension of M-theory, 
as we shall see in detail in the next Section.

In heterotic models, the ten-dimensional string coupling 
$\lambda_H$ and the string scale $M_H = (\alpha')^{-1/2}$ are expressed in terms of four
dimensional parameters as
\be
\lambda_H= \frac{(\alpha_G)^2}{8 \sqrt{2}} \ V^{1/2} M_P^3 \ , \qquad  M_H=({\alpha_G
\over 8})^{1/2} M_P \ , \label{U2}
\ee where $(2\pi)^6 V$ is the volume of the six-dimensional internal manifold,
$\alpha_G \sim {1 \over 25}$ is the gauge coupling at the unification  scale and
$M_P=G_N^{- 1/2}$ is the Planck mass. The corresponding relations for the  open
superstring can be deduced using the duality between the heterotic SO(32) and the type
I string in 10d \cite{W1}:
\be
\lambda_I={1 \over \lambda_H} \ , \qquad
\ M_I=M_H {\lambda_H}^{-1/2}
 \ . \label{U3}
\ee Thus
\be
\lambda_I=\frac{8 \sqrt{2}}{(\alpha_G)^2} \ V^{-1/2} M_P^{-3} \ , 
\qquad M_I=({\sqrt 2 \over \alpha_G M_P})^{1/2} V^{-1/4} \quad , 
\label{U4}
\ee where $\alpha_G$ is the coupling for nine-brane gauge fields. The unification scale
$M_{GUT}$ is fixed by the fundamental mass scale
$M_I$, and therefore in the following we take $M_I \sim 3 \times  10^{16} GeV$.

In the simplest case of an isotropic compact space with $V=r^6 M_I^{-6}$,  the
natural values of the dimensionless radius $r$ are of order one.  In particular, $r=1$
corresponds to the self-dual point of circle compactification.  In this case
$M_I=({\alpha_G / \sqrt 2}) r^3 M_P$, that requires
$r \sim{1 \over 6}$, and all  the corresponding radii are of the order of the
unification scale. 

The situation changes drastically for anisotropic compactifications. For instance, if
one radius $R_I$ is large and five others are of order
$r M_I^{-1}$ ($V=r^5 M_I^{-5} R_I$), eq. (\ref{U4}) leads to
\be R_I^{-1}=({\alpha_G \over \sqrt 2})^2 r^5 {M_P^2 \over M_I} \ , \qquad
\lambda_I = {8 \over {\alpha_G}} ({M_I \over M_P})^2  \ . \label{U6}
\ee Since $R_I^{-1} >>M_I$, it is convenient to perform a T-duality along the $R_I$
direction to turn to the type I$^\prime$ description. The corresponding duality
relations are
\be  M_{I^\prime}=M_I \ , \qquad R_{I^\prime}=M_I^{-2}/R_I \ , \qquad 
\lambda_{I^\prime} =\lambda_I/(R_IM_I) \quad , \label{U65}
\ee and therefore
\ba R_{I^\prime}^{-1}=({\sqrt 2 \over \alpha_G})^2 r^{-5} {M_I^3 \over M_P^2} \ , 
\qquad \lambda_{I^\prime}=4 \alpha_G  r^{5}  \ . \label{U7}
\ea        
Interestingly, these relations are similar to those obtained by  Horava and Witten, if
$R_{I^\prime}$ is identified with the radius of the eleventh dimension 
in M theory \cite{HW,AQ}. The
relation between the corresponding BPS spectra will be discussed in detail in the
following Section. Actually, eq. (\ref{U7}) motivated the suggestion \cite{AQ} that the
Scherk-Schwarz mechanism along the eleventh dimension of M-theory  can describe gaugino
condensation with a condensation scale 
$\Lambda \sim M_I$, since in the latter case $m_{3/2} \sim \Lambda^3/M_P^2$. In
particular, with $r
\sim 1$ eq. (\ref{U7}) gives 
$m_{3/2} \sim R_{I^\prime}^{-1} \sim 10^{13}$ GeV.   In the following Sections we will
see that, in a class of models,  charged matter fields associated to branes orthogonal
to the dimension of radius
$R_{I^\prime}$  feel the breaking of supersymmetry only through Planck suppressed
radiative corrections. Therefore, this scenario is of potential interest for
phenomenology. 

If two radii $R_I$ are large and the other four have natural values $rM_I^{-1}$, one
finds
\ba R_I^{-1} &=& {\alpha_G r^2 \over \sqrt 2} M_P \ , \qquad \lambda_I =  {8 \over 
\alpha_G} ({M_I \over M_P})^2
\ , \nonumber \\ R_{I^\prime}^{-1} &=& {\sqrt 2 \over \alpha_G r^2} {M_I^2 \over M_P} \
, \qquad
\lambda_{I^\prime}=4 \alpha_G r^4  \ . \label{U8}
\ea 
Once more, it is convenient to use the type I$^\prime$ description, but now
$R_{I^\prime}^{-1}$ is raised by roughly one order of magnitude. While this may still be
acceptable for phenomenology, additional large internal radii have the effect of moving
the intermediate scale further toward the unification scale.

\section{BPS spectra in nine dimensions}

The basic features of the breaking mechanism are already visible in the compactification
from ten to nine dimensions. Therefore, in this Section we restrict our attention to 
the type I theory compactified on a circle of radius $R$. T-duality relates this
analysis to the Horava-Witten reduction of M theory, and thus provides an explicit
description of the  Scherk-Schwarz breaking along the eleventh dimension. The
compactification of M-theory on $S^1 \times S^1/Z_2$ (with radii
$R_{10}$ and $R_{11}$, respectively) admits two different interpretations \cite{HW}:
\begin{itemize}
\item[1. ]as M-theory on $S^1/Z_2\times S^1$, that according to \cite{HW}  describes
the $E_8
\times E_8$ heterotic string of coupling $\lambda_{E_8
\times E_8}\!=\!(R_{11}M_{11})^{3/2}$, compactified on a circle $ S^1$ of radius
$R_{10}(R_{11}M_{11}\!)^{1/2}$. In this case, a Wilson line must be added, and the
theory is in a vacuum with an unbroken $SO(16) \times SO(16)$ gauge group.
\item[2. ]as M-theory on $S^1 \times S^1/Z_2$, that according to \cite{W1}  describes
the IIA theory of coupling $\lambda_{IIA} = (R_{10}M_{11})^{3/2}$, compactified  further
on the
$S^1/Z_2$ orientifold of radius $R_{11}(R_{10}M_{11})^{1/2}$. The result is the
type-I$^\prime$ theory,  T-dual (with respect to the eleventh coordinate) to the type I
theory (in its $SO(16)
\times SO(16)$ vacuum), with coupling
$\lambda_I = R_{10}/R_{11}$, compactified on a circle of radius $M_I^{-2} /(R_{11}
(R_{10}M_{11})^{1/2})$. In the M-theory regime ($R_{11} >> R_{10}$), the type I and type
I$^\prime$ theories can both be weakly coupled, and can consequently be treated as
perturbative strings.
\end{itemize}

It is particularly instructive to translate in type I or heterotic language
the masses of the BPS states of M-theory \cite{HW}. 
Consider first the Kaluza-Klein states of the supergravity multiplet on
$T^2=S^1\times S^1$,  together with the wrapping modes of the membrane around the 
torus. Their masses are 
\ba {\cal M}^2 = {l^2 \over R_{11}^2} + {m^2 \over R_{10}^2} +  n^2 R_{10}^2 R_{11}^2
M_{11}^6 \quad ,
\label{N1}
\ea where $( l,m,n )$ is a triplet of integers labelling the corresponding charges.  In
the effective field theory, supersymmetry breaking along the eleventh dimension results
from shifts \cite{SS}
\be 
l \rightarrow l + \omega \quad , \label{lshift}
\ee 
where $\omega$ is the R-charge of the corresponding state. Ordinary Kaluza-Klein
excitations in the tenth direction with charges $m$ do not feel the breaking, while the
wrapping modes labelled by
$n$ in principle do.

In type-I and type-I$^\prime$ units, the masses of the states (\ref{N1}) are
\ba 
{\cal M}_I^2 = l^2 R_I^2 M_I^4+ {m^2 R_I^2 M_I^4 \over {\lambda_I}^2}  + {n^2 \over
R_I^2} \ ,
\qquad {\cal M}_{I^\prime}^2 = {l^2 \over R_{I^\prime}^2} +  {m^2 M_I^2 \over
{\lambda_{I^\prime}}^2} + n^2 R_{I^\prime}^2 M_I^4  \ . \label{N2}
\ea 
Therefore, from the type I$^\prime$ viewpoint the breaking along the eleventh 
dimension gives Kaluza-Klein type masses proportional to 
$R_{I^\prime}^{-1}$, and supersymmetry is restored in the $R_{I^\prime}
\rightarrow \infty$ limit.  This is the string generalization of the field-theoretical
Scherk-Schwarz mechanism. In the zero-winding sector
$(n=0)$, it reduces to the ordinary Scherk-Schwarz breaking effected by the momentum
shift (\ref{lshift}), while the extension to the whole (perturbative) closed-string 
spectrum is
determined by modular invariance. On the other hand, from the type I viewpoint, this
breaking gives winding type masses
proportional to $R_I M_I^2$, and supersymmetry is restored in the 
$R_I \rightarrow 0$ limit. As a result, there is no field-theoretical description
corresponding to this case.

In a similar fashion, in 
$E_8 \times E_8$ and $SO(32)$ heterotic units the states (\ref{N1}) have masses
\ba 
{\cal M}_{E_8}^2 = {l^2 M_H^2 \over {\lambda_{E_8}}^2} + {m^2 \over R_{E_8}^2} + n^2
R_{E_8}^2 M_H^4 \ , \quad {\cal M}_H^2 = l^2 {R_H^2 M_H^4 \over {\lambda_H}^2} + m^2
R_H^2 M_H^4 +  {n^2 \over R_H^2} \quad .
\label{N3}
\ea 
Notice that in both heterotic theories the shift $l \rightarrow l+ \omega$  gives 
masses of non-perturbative type, and thus the breaking along the eleventh dimension is a
genuinely non-perturbative phenomenon. Similarly, the perturbative heterotic
states labelled by $m$ turn into non-perturbative ones in the type I or type I$^\prime$
descriptions.       
  
There are also twisted M-theory states associated to the fixed points of 
$S^1/Z_2$, that are charged under the gauge group. 
These include ordinary momentum excitations in the tenth direction and
membrane wrappings in the full internal space, for which 
\be {\cal M}^2={{\tilde m}^2 \over R_{10}^2} + {\tilde n}^2 R_{10}^2 R_{11}^2  M_{11}^6
\quad . 
\label{N4}
\ee 
As before, the charged Kaluza-Klein states labelled by ${\tilde m}$ do not feel the
Scherk-Schwarz breaking along the eleventh coordinate, 
but the wrapping modes ${\tilde n}$ in principle
do. In type I and type I$^\prime$ units, their masses become
\ba 
{\cal M}_I^2 = {{\tilde m}^2 R_I^2 M_I^4 \over {\lambda_I}^2} +  {{\tilde n}^2 \over
R_I^2} \ ,
\qquad {\cal M}_{I^\prime}^2 =  {{\tilde m}^2 M_I^2 \over {\lambda_{I^\prime}^2}} +
{\tilde n}^2 R_{I^\prime}^2 M_I^4 \quad , \label{N5}
\ea 
and the wrapping modes are thus perturbative open string states. In $E_8 \times E_8$
and $SO(32)$ heterotic units, the masses of the charged states are
\ba 
{\cal M}_{E_8}^2 = {{\tilde m}^2 \over R_{E_8}^2} + {{\tilde n}^2 R_{E_8}^2}M_H^4 \ ,
\qquad {\cal M}_H^2 = {\tilde m}^2 R_H^2 M_H^4 + {{\tilde n}^2 \over R_H^2} \ . 
\label{N6}
\ea 
{}From the $E_8 \times E_8$ viewpoint, the breaking along the eleventh dimension  can
affect the windings, that can be seen only at the string level. 
Thus, from the field
theory (Kaluza-Klein) point of view, the charged states are unaffected.

This mechanism can be contrasted with the ordinary Scherk-Schwarz mechanism of
perturbative heterotic strings, that amounts to shifts of $n$ and $\tilde n$ (or
$m$ and $\tilde m$) for the $SO(32)$ (or $E_8\times E_8$) model (see eqs. (\ref{N3}) and
(\ref{N6})). The perturbative states labelled by 
$n$ and $\tilde n$ have counterparts in the type
I theory that reflect the perturbative nine-dimensional duality between the two models
(see eqs. (\ref{N2}) and (\ref{N5})). This is effective if the string coupling
$\lambda_I$ is small and $R_I$ is large.

\section{Scherk-Schwarz mechanism for oriented closed strings}

In order to make the previous analysis more explicit, in this Section we review the
basic features of the  breaking mechanism in models of oriented closed strings, following
\cite{KP}. In the next Section we shall extend these results to the open  descendants,
describing corresponding breaking patterns in type I (or  type I$^\prime$) models. The
Scherk-Schwarz mechanism results from a discrete deformation of the type-IIB partition
function compatible with modular invariance \cite{r,KP}. 
The starting point of our discussion is thus  the partition function for the circle
reduction of the type-IIB string,
\be 
{\cal T}= {\tau_2^{-7/2} \over |\eta(\tau)|^{22}}
\sum_{m,n=-\infty}^{\infty} Z_{m,n}(\tau,{\bar \tau}) {\left| \sum_{\rm spin \atop
structures} C
\left( 
\begin{array}{c}  {\bf a} \\ {\bf b} 
\end{array} 
\right)
  \theta^4
\left(
\begin{array}{c} {\bf a} \\ {\bf b}
\end{array}
\right) (\tau) \right|}^2 \ , \label{N7}
\ee 
where $(m,n)$ is a pair of integers labelling the Kaluza-Klein momentum and the
winding number in the compactified dimension, and the $C$'s are phases 
specifying the GSO
projection and depending on the spin structures labelled by the pair 
$({\bf a}, {\bf b})$. 
The $\theta$'s are Jacobi theta functions, $\eta$ is the Dedekind function,
$\tau=\tau_1+i\tau_2$ is the modular parameter of the world-sheet torus,
and\footnote{For brevity, in this Section and in the following we set $\alpha'=2$.}
\be
\sum_{m,n} Z_{m,n}(\tau,{\bar \tau})= {1 \over |\eta(\tau)|^2}
\sum_{m,n} q^{1/2({m \over R}+{nR \over 2})^2} {\bar q}^{1/2({m \over R}-{nR \over
2})^2}= {R{\tau_2}^{-1/2} \over{\sqrt 2} |\eta(\tau)|^2} \sum_{{\tilde m},n} e^{-{\pi
R^2 \over 2 \tau_2} |{\tilde m}+n \tau|^2 } \ , \label{N8}
\ee
where $q=e^{2i\pi \tau}$. The last form of this expression, obtained by a Poisson
resummation in
$m$, will be used below to perform the Scherk-Schwarz deformation in a way that keeps
modular invariance manifest. On the other hand,  the hamiltonian formulation gives the
whole expression a lattice interpretation. 
In this case the partition function takes the form
\be 
Z (\tau, {\bar \tau})= {\rm Tr}\ q^{L_0}\ {\bar q}^{\bar L_0} \ ,
\label{N9}
\ee where
\be L_0={1 \over 2}{\bf p}_L^2+{1 \over 2} (p_L^0)^2 -{1 \over 2} + \dots \ , \quad
{\bar L}_0={1
\over 2}{\bf p}_R^2+{1 \over 2} (p_R^0)^2 -{1 \over 2} + \dots 
\label{N10}
\ee  
Here the dots stand for the contributions of string oscillators, while
$p_{L,R}^0=m/R \pm nR/2$ are the left and right momenta associated to the
$\Gamma_{(1,1)}$ lattice of the circle. Moreover, we use bosonized fermionic coordinates
that give rise to an additional lattice $\Gamma_{(4,4)}$ with momenta ${\bf p}_{L,R}$.

The general method for breaking supersymmetry by compactification uses a (discrete)
R-symmetry of the higher-dimensional theory and couples the lattice momenta to the
corresponding
R-charges ${\bf e}$. The R-transformations are in general discrete remnants of
internal rotations. For simplicity, we shall restrict our attention to the $Z_2$ case,
that corresponds to a
$2\pi$-rotation and therefore affects only spinorial representations. Moreover, this is
the only possibility in nine dimensions, where it reduces to fermion number parity. 
Upon
bosonization of the fermionic coordinates, as described above, this construction
translates into a Lorentz boost that mixes the
$\Gamma_{(1,1)}$ and $\Gamma_{(4,4)}$ lattices consistently with modular invariance:
\ba 
&{\bf p}_L \rightarrow {\bf p}_L-{\bf \mbox{\boldmath $\omega$}}_L (p_L^0-p_R^0) \ , 
\quad &{\bf p}_R \rightarrow {\bf p}_R-{\bf \mbox{\boldmath $\omega$}}_R (p_L^0-p_R^0) \
, 
\nonumber \\ &{p_L^0} \rightarrow  p_L^0 + \omega \cdot p - {1 \over 2}
\omega\cdot\omega (p_L^0-p_R^0) \ , \quad &{p_R^0} \rightarrow  p_R^0 + \omega \cdot p -
{1 \over 2}
\omega\cdot\omega (p_L^0-p_R^0) \  , \label{N11}
\ea 
where ${\bf \mbox{\boldmath $\omega$}}  = {{\bf e} / R}$ \cite{KP}. As usual, the
scalar products in (\ref{N11}) are defined with a lorentzian metric, so that for
instance
$\omega \cdot p = {\bf \mbox{\boldmath $\omega$}}_L \cdot {\bf p}_L-{\bf \mbox{\boldmath
$\omega$}}_R \cdot {\bf p}_R$. The boost (\ref{N11}) translates into the shift:
\be 
m \rightarrow m + e\cdot p - {n \over 2} e\cdot e \ , \ n \rightarrow n
\ , \qquad {\bf p} \rightarrow {\bf p} - n {\bf e} \ , \label{N12} 
\ee 
that in the zero winding sector clearly reproduces the standard
Scherk-Schwarz momentum shifts of the effective field theory. In this work we are  
particularly interested in left-right symmetric charge assignments, since these are the
starting point for the construction of type-I descendants. In this case, the above  
equations simplify, since
${\bf e}_L={\bf e}_R$, and therefore $e\cdot e=0$.

The deformed partition function takes a rather compact form in the lagrangian
formulation obtained by the Poisson resummation (\ref{N8}):
\ba 
{\cal T} &=& {R\tau_2^{-4} \over{\sqrt 2} |\eta(\tau)|^{24}}
\sum_{n,{\tilde m}=-\infty}^{\infty} e^{-{\pi R^2 \over 2\tau_2} |{\tilde m} + n
\tau|^2} \sum_{\rm spin \atop structures} {\tilde C}
\left(
\begin{array}{c} {\bf a}_L \\ {\bf b}_L
\end{array}
\right) {\tilde C}^*
\left(
\begin{array}{c} {\bf a}_R \\ {\bf b}_R
\end{array}
\right)\times
\nonumber \\ &&{\prod_{i_L=1}^4} \theta
\left(
\begin{array}{c} {\bf a}_{i_L}-n{\bf e}_{i_L} \\ {\bf b}_{i_L}+{\tilde m} {\bf e}_{i_L}
\end{array}
\right) (\tau) \ 
\prod_{i_R=1}^4 {\bar \theta}
\left(
\begin{array}{c} {\bf a}_{i_R}-n{\bf e}_{i_R} \\ {\bf b}_{i_R}+{\tilde m} {\bf e}_{i_R}
\end{array}
\right) ({\bar \tau}) \ , \label{N16}
\ea 
where
\be {\tilde C}
\left(
\begin{array}{c} {\bf a} \\ {\bf b}
\end{array}
\right) = e^{2 i \pi n {\bf e} ( {\bf b}+ { {\tilde m} \over 2} {\bf e})} C
\left(
\begin{array}{c} {\bf a} \\ {\bf b}
\end{array} \right)  . \label{N17}
\ee 
It follows from the partition function, or equivalently from the shifts (\ref{N12}),
that in nine dimensions this breaking yields mass shifts for all fermions. In
particular,
the gravitino and all previously massless fermions acquire a mass equal to $e\cdot p/R$.

In the next Section we will extend these results to unoriented closed and open strings
\footnote{An alternative way to break supersymmetry using internal magnetic
fields is described in \cite{Ba}.}.
The standard construction of open descendants \cite{S,PS,BS} 
then associates to the unoriented sector
of the type IIB theory of eq. (\ref{N16}) an open sector where supersymmetry is broken
by shifts as in eq. (\ref{N12}) for all fermions, although with $n=0$. This is
effectively the field theory realization of the Scherk-Schwarz mechanism. 
An even more interesting possibility is realized if the open strings have Dirichlet
boundary conditions along the circle. This is the case for type-I$^\prime$ theory, where
the circle is identified with the eleventh coordinate of M theory.  For the sake of
comparison with the previous, standard case, one can perform a T-duality to revert to
the type-I description. In this case the shifts introduced in the closed sector 
affect the
windings, rather than the momenta. Thus, one would expect that the open spectrum be
untouched. The explicit analysis of the next Section will show that, while the actual
situation is more involved, this naive expectation is realized for the massless
modes. The relevant boost for the type IIB theory can be obtained from eq. (\ref{N11})
by a T-duality. The result is
\ba 
&{\bf p}_L \rightarrow {\bf p}_L-{\bf \mbox{\boldmath $\eta_L$}} (p_L^0+p_R^0) \ ,
\quad &{\bf p}_R \rightarrow {\bf p}_R-{\bf \mbox{\boldmath $\eta_R$}} (p_L^0+p_R^0) \ ,
\nonumber \\ &p_L^0 \rightarrow  p_L^0+ \eta\cdot p-{1 \over 2}
\eta\cdot\eta (p_L^0+p_R^0) \ , \quad &p_R^0 \rightarrow  p_R^0 -\eta\cdot p + {1 \over
2}
\eta\cdot\eta (p_L^0+p_R^0) \ , \label{N14}
\ea where ${\bf \mbox{\boldmath $\eta$}}  = {\bf e}R/2$. The boost (\ref{N14})
corresponds to the shift of the lattice momenta
\be m \rightarrow m \ , \ n \rightarrow n + e\cdot p - {m \over 2} e\cdot e 
\ , {\bf p} \rightarrow {\bf p}-m {\bf e}\ , \label{N15}
\ee 
clearly related to eq. (\ref{N12}) by the interchange of $m$ and $n$. The gravitino
mass is now
$e\cdot pR/2$, and supersymmetry is thus recovered in the $R\to 0$ limit. Finally, the
explicit form of the partition function can be obtained from eq. (\ref{N16}) replacing
$n$ with $m$ and $\tilde m$ with $\tilde n$, where ``tilde" indicates Poisson resummed
indices. Again, the type-I descendants are associated to left-right symmetric choices of
charges ${\bf e}_L={\bf e}_R$, so that $e\cdot e=0$.

One can actually write the partition function (\ref{N16}) and the corresponding one for
the shifts (\ref{N15}) in a form 
more suitable for building the open descendants. To this end, let us define the level
one $SO(2n)$ characters 
\ba 
I_{2n} &=& {1 \over 2 \eta^n} ( \theta_3^n + \theta_4^n) \ , \qquad\quad 
V_{2n}={1 \over 2 \eta^n} (
\theta_3^n - \theta_4^n) \ , \nonumber \\ S_{2n} &=& {1 \over 2 \eta^n} ( \theta_2^n +
i^n
\theta_1^n) \ , \qquad C_{2n}={1 \over 2 \eta^n} ( \theta_2^n - i^n \theta_1^n) \ ,
\label{E1}
\ea 
where $\theta_i$ are the four Jacobi theta-functions with integer
characteristics. Then, leaving aside all contributions from transverse bosons and from
the measure over the moduli, the partition function of the type IIB superstring in 10d
is simply
\be 
T_{\rm IIB}=|V_8-S_8|^2 \ . \label{E2}
\ee 
In this notation, the massless spectrum is manifest; in particular,
$|V_8|^2$ describes the universal bosonic modes of the NS-NS sector (graviton, dilaton
and 2-index antisymmetric tensor), while $|S_8|^2$
describes the additional scalar, the 2-form and the self-dual 4-form of the R-R sector. 

In a similar fashion, after a Poisson resummation back to the hamiltonian form, the
deformed partition function (\ref{N16}) for the ordinary Scherk-Schwarz breaking in nine
dimensions with
${\bf e}_L={\bf e}_R=(0,0,0,1)$ becomes
\be 
{\cal T}_1 = E^{\prime}_0 ( V_8 {\bar V}_8 + S_8 {\bar S}_8 ) +  O^{\prime}_0 ( I_8
{\bar I}_8 + C_8 {\bar C}_8 ) - E^{\prime}_{1/2} ( V_8 {\bar S}_8 + S_8 {\bar V}_8 ) - 
O^{\prime}_{1/2} ( I_8 {\bar C}_8 + C_8 {\bar I}_8 ) \quad . \label{E3}
\ee 
Here, for brevity, we have introduced the projected lattice sums
\ba 
&&E^{\prime}_0 = \sum_{m,n} {1 + (-1)^n \over 2} Z_{m,n} \ ,\qquad\qquad
O^{\prime}_0 =
\sum_{m,n} {1 - (-1)^n \over 2} Z_{m,n} \ , \nonumber \\ &&E^{\prime}_{1/2} = \sum_{m,n}
{1 + (-1)^n
\over 2} Z_{m+1/2,n} \ ,\qquad O^{\prime}_{1/2} = \sum_{m,n} {1 - (-1)^n \over 2}
Z_{m+1/2,n} \quad ,
\label{E4}
\ea 
where $E^\prime$ and $O^\prime$ refer to even and odd windings and the subscripts
$0$ and $1/2$ refer to unshifted and shifted momenta. In this string generalization of
the Scherk-Schwarz mechanism, all space-time fermions have evidently masses
shifted by
${1/(2R)}$ compared  to the supersymmetric case. Modular invariance, however, changes the
GSO projection  in the odd winding sector. As a result, for $R\le\sqrt{\alpha^\prime}$
the spectrum contains a
tachyon associated to $I_8 {\bar I}_8$.\footnote{In the thermal case,
$T=1/2\pi\sqrt{\alpha^\prime}$ corresponds to the value of the Hagedorn temperature for
the type II theory.} 
The supersymmetric type IIB model (\ref{E2}) in 10d is recovered in
the $R \rightarrow\infty$ limit.

Alternatively, this model can be obtained as an asymmetric orbifold of the
non-super\-symmetric 0B model with partition function
\be 
T_{\rm 0B}=|I_8|^2+|V_8|^2+|S_8|^2+|C_8|^2 \ , \label{E200}
\ee 
using the discrete symmetry $g=-(-1)^{G_L} (-1)^n$, where
$G_L$  is the world-sheet left fermion number, so that $-(-1)^{G_L}$ acts as $1$  on
$V_8,S_8$ and
$-1$ on $I_8,C_8$. This model can also be described as a symmetric orbifold of the IIB
superstring of eq. (\ref{E2}) by the $Z_2$ symmetry $(-1)^F I$, where
$F=F_L+F_R$ is the space-time fermion number and the shift $I:X_9 \rightarrow X_9 + \pi
R$ acts on states as $(-1)^m$. The resulting partition function ${\cal T}_1^\prime$
coincides with ${\cal T}_1$ upon doubling the radius, so that ${\cal T}_1(R)={\cal
T}_1^\prime(2R)$. More explicitly 
\be 
{\cal T}_1^\prime = E_0 ( V_8 {\bar V}_8 + S_8 {\bar S}_8 ) + E_{1/2}  ( I_8 {\bar
I}_8 + C_8 {\bar C}_8 ) - O_0 ( V_8 {\bar S}_8 + S_8 {\bar V}_8 ) - O_{1/2}  ( I_8 {\bar
C}_8 + C_8 {\bar I}_8 ) \ ,\label{E5} 
\ee 
where $E_0$, $O_0$, $E_{1/2}$, $O_{1/2}$ are defined as in eq. (\ref{E4}) but with
$m$ and $n$ interchanged.

Finally, the type IIB deformed model with windings shifted according to eq. (\ref{N15})
can be obtained in a straightforward way from eq. (\ref{E3}) interchanging momenta and
windings. The resulting partition function is
\be 
{\cal T}_2 = E_0 ( V_8 {\bar V}_8 + S_8 {\bar S}_8 ) + O_0 ( I_8 {\bar I}_8 + C_8
{\bar C}_8 ) - E_{1/2} ( V_8 {\bar S}_8 + S_8 {\bar V}_8 ) - O_{1/2} ( I_8  {\bar C}_8 +
C_8 {\bar I}_8 ) \ .
\label{E19}
\ee 
This model is tachyon-free for values of the radius $R \le\sqrt{\alpha^\prime}$,
while the supersymmetric 10d IIB theory is recovered in the $R \rightarrow 0$ limit. A
field theory interpretation is thus more natural in the T-dual type-IIA description,
where the shifts are transferred back to the momenta. The corresponding open sector,
however, corresponds to the type I$^\prime$ construction with Dirichlet boundary
conditions. Since the resulting open strings have only windings, the shifted momenta are
naturally associated to the eleventh direction of M theory, 
as we have shown in Section 3.
In the following, we shall refer to this model as the M-theory breaking model. Note that
it can also be obtained as an asymmetric orbifold of the 0B theory (\ref{E200}), using
the discrete symmetry $g=-(-1)^{G_L} I$.

\section{Explicit type I models in nine dimensions}
\vskip 10pt
\begin{flushleft} {\large\bf 5.1. \ Review of the construction procedure}
\end{flushleft}
\vskip 10pt
\noindent 
Before describing the nine-dimensional type I models, we would like to present
a brief review of the algorithm that we use. This was introduced in \cite{PS,BS}, and
developed further in \cite{PSS}. Type I models can be obtained as ``orbifolds" of
left-right symmetric type-IIB models by the world-sheet
involution $\Omega$ that interchanges left and right movers \cite{S}.
The starting point thus consists in
adding to the (halved) torus amplitude the Klein-bottle
${\cal K}$.  This completes the projection induced by $\Omega$, 
and is a linear combination
of the diagonal contributions to the torus amplitude, with argument $q {\bar q}$, where
$q$ was defined after eq. (\ref{N7}). Thus, starting from
\be {\cal T}=\sum_{ij}\, X_{ij}\, \chi_i(\tau)\, \chi_j(\taubar )\quad ,
\label{T}
\ee where the $\chi$'s are a set of characters of the underlying conformal field theory
and $X$ is a matrix of integers, one obtains \footnote{As discussed in \cite{PSS}, 
in general one
has the option to modify eq. (\ref{K}), altering $X_{ii}$ by signs $\epsilon_i$. These
turn sectors symmetrized under left-right interchange into antisymmetrized ones,
and vice-versa, and are in general constrained by compatibility with the fusion rules.}
\be 
{\cal K}= \frac{1}{2} \ \sum_{i}\, X_{ii} \, \chi_i(2i\tau_2) \ ,
\label{K}
\ee 
with $\tau_2$ the proper time for the closed string. In order to identify the
corresponding open sector, it is useful to perform the $S$ modular transformation
induced by 
\ba 
{\cal K}:\qquad\qquad 2\tau_2 \ &{{{}\atop\longrightarrow}\atop {{}\atop S}}& 
\ {1\over 2\tau_2}\equiv l\ ,
\label{tl}
\ea
thus turning the direct-channel Klein-bottle amplitude
${\cal K}$ into the transverse-channel amplitude $\tilde{\cal K}$. The latter describes
the propagation of the closed spectrum on a tube of length $l$ terminating at two
crosscaps\footnote{The crosscap, or real projective plane, is a non-orientable surface
that may be defined starting from a 2-sphere and identifying antipodal points.}, and
has the generic form
\be 
{\tilde{\cal K}}= \ \sum_{i}\, \Gamma_i^2 \, \chi_i(il) \ ,
\label{Ktilde}
\ee 
where the coefficients $\Gamma_i$ can be related to the one-point functions of the
closed-string fields of ${\cal T}$ in the presence of a crosscap.

The open strings correspond to the twisted sector of the spectrum with respect to the
involution
$\Omega$, and may be deduced from the closed-string spectrum in a similar fashion.
First, the direct-channel annulus amplitude ${\cal A}$
may be deduced from the transverse-channel boundary-to-boundary amplitude $\tilde{\cal
A}$. This has the general form \cite{BS}
\be 
{\tilde{\cal A}}= \ \sum_{i}\, B_i^2 \, \chi_i(il) \ ,
\label{Atilde}
\ee 
where the coefficients $B_i$ can be related to the one-point functions of
closed-string fields in the presence of boundaries. The relevant $S$ modular
transformation now maps the closed string proper time $l$ on the tube into the
open-string proper time $t$ on the annulus according to
\ba 
{\cal A}:\qquad\qquad l \ &{{{}\atop\longrightarrow}\atop {{}\atop S}}&  
{1\over l}\equiv {t_{\cal A}\over 2} \ .
\label{tlA}
\ea
The direct-channel annulus amplitude then takes the form
\be 
{\cal A}= \ \frac{1}{2} \ \sum_{i,a,b} \ A^i_{ab} \ n_a\ n_b\ 
\chi_i\left({it_{\cal A} \over 2}\right)\ ,
\label{A}
\ee 
where the $n$'s are integers that have the interpretation of Chan-Paton
multiplicities for the boundaries and the $A^i$ are a set of matrices with integer
elements. These matrices are obtained solving diophantine equations determined by the
condition that the modular transform of eq. (\ref{Atilde}) involve only integer
coefficients, while the Chan-Paton multiplicities arise as free parameters of the
solution.

Finally, the transverse-channel M{\"o}bius amplitude $\tilde{\cal M}$ describes the
propagation of closed strings between a boundary and a crosscap, and is determined by
factorization from
$\tilde{\cal K}$ and $\tilde{\cal A}$. It contains the characters common to the two
expressions with coefficients that are geometric means of those present in
$\tilde{\cal K}$ and $\tilde{\cal A}$ \cite{BS}. Thus
\be 
{\tilde{\cal M}}= \ 2 \sum_{i}\, B_i\ \Gamma_i \ {\hat\chi}_i\ (il+\frac{1}{2}) \ ,
\label{Mtilde}
\ee 
where the hatted characters form a real basis and are obtained by the redefinitions
\be 
{\hat\chi}_i\ (il+\frac{1}{2})=e^{-i\pi h_i}\chi_i\ (il+\frac{1}{2})\ ,
\label{hat}
\ee 
where $h_i$ are the conformal weights of the corresponding primary fields. 
The direct-channel
M{\"o}bius amplitude ${\cal M}$ can then be related to $\tilde{\cal M}$ by a
modular $P$ transformation and a redefinition
\ba
{\cal M}:\qquad\qquad {it_{\cal M} \over 2}+1/2 \ 
&{{{}\atop\longrightarrow}\atop {{}\atop P}}&  
{i \over 2t_{\cal M}}+{1 \over 2} \equiv il + {1 \over 2} \ \label{hatt}
\ea
This is realized on the hatted characters by the sequence
$P=T^{1/2} ST^2ST^{1/2}$, with $T$ the diagonal 
matrix that implements the transformation
$\tau\to\tau+1$. The direct-channel M{\"o}bius amplitude then takes the form
\be 
{\cal M}= \ \frac{1}{2} \ \sum_{i,a} \ M^i_{a} \ n_a\  {\hat\chi}_i\left({it_{\cal M}
\over 2}+\frac{1}{2}\right)\ ,
\label{M}
\ee 
where by consistency the integer coefficients $M^i_{a}$ satisfy the constraints
\be 
M^i_{a}=A^i_{aa} \qquad ({\rm mod 2})\ ,
\label{Mia}
\ee
that make $\cal M$ the completion of $\cal A$.
The full one-loop vacuum amplitude is then
\be 
\int\left({1\over 2}{\cal T}(\tau,\taubar)+{\cal K}(2i\tau_2)
+{\cal A}({it\over 2})+{\cal M}({it\over 2}+{1\over 2})\right)\ ,
\label{typeI}
\ee 
where the different measures of integration are left implicit.
In the remainder of this paper, we shall often omit the dependence on 
world-sheet modular parameters.

For the models of interest, in order to link the direct and transverse channels, one
needs
the transformation matrices $S$ and $P$ for the level-one $SO(2n)$ characters of eq.
(\ref{E1}). These may be simply deduced from the corresponding transformation properties
of the Jacobi theta functions, and are
\be  
S_{(2n)} =  {1 \over 2}
\left(
\begin{array}{cccc}  1 & 1 & 1 & 1 \\ 1 & 1 & -1 & -1 \\ 1 & -1 & i^{-n} & -i^{-n} \\ 1
& -1 & -i^{-n} & i^{-n}
\end{array}
\right) \ , \  P_{(2n)} = 
\left(
\begin{array}{cccc}  c & s & 0 & 0 \\  s & -c & 0 & 0 \\ 0 & 0 & \zeta c & i \zeta s \\
0 & 0 & i
\zeta s & \zeta c
\end{array}
\right) \ ,\label{E201}
\ee   where $c= \cos {n \pi /4}$, $s= \sin {n \pi /4}$ and $\zeta= e^{-i{n \pi/4}}$. 

For later use, let us also define
\be 
Z_{m+a}(\tau) = \frac{q^{{1\over 2} {\left(\frac{m+a}{R}\right)}^2}}{\eta(\tau)}\ ,
\qquad
\tilde{Z}_{n+ b}(\tau) =\frac{q^{{1\over 2} {\left( (n + b){R\over 2}
\right)}^2}}{\eta(\tau)} \ . \label{Edef}
\ee 
In relating the direct and transverse channels, one also needs the Poisson
transformation
\be
\sum_m e^{2i\pi m b} Z_{m+a} (-{1 \over \tau})=R\ e^{-2i\pi ab} \sum_n e^{-2i
\pi na} {\tilde Z}_{2n+2b} ({\tau}) \ . \label{E60}
\ee 

\vskip 10pt
\begin{flushleft} {\large\bf 5.2. \ Scherk-Schwarz breaking model}
\end{flushleft}
\vskip 10pt
\noindent We can now proceed to construct the open descendants of the two deformed type
IIB models described by ${\cal T}_1$ and ${\cal T}_2$ of eqs. (\ref{E3}) and
(\ref{E19}). To this end, we begin by applying the $\Omega$ projection, that symmetrizes
the NS-NS sector and antisymmetrizes the R-R sector. The Klein bottle contribution to
the partition function completes the projection of the closed sector, and thus receives
contributions from all sectors invariant under $\Omega$. These include, in particular,
the sublattice
\be {\bf p}_L={\bf p}_R \ , \ p_L^0=p_R^0 \ . \label{N13}
\ee It is then clear that for the Scherk-Schwarz breaking model (\ref{E3}) the Klein
bottle is  unaffected by supersymmetry breaking. Indeed, the shifts of eq. (\ref{N12})
vanish identically, since $e\cdot p=e\cdot e=0$ for the states satisfying the conditions
(\ref{N13}). The resulting Klein bottle projection ${\cal K}_1$ is
\be  
{\cal K}_1 = \frac{1}{2}  \ (V_8 - S_8) \ \sum_m\ Z_m
\ , \label{E7}
\ee 
where we have left implicit an overall factor 
$\tau_2^{-11/2}\eta^{-7}\ d\tau_2$. The
transverse-channel Klein bottle amplitude is then
\be
\tilde{\cal K}_1 = \frac{2^{9/2}}{2} ( V_8 - S_8 )  R\sum_n\ \tilde{Z}_{2n} \ ,
\label{E8}
\ee 
where the numerical factor $2^{9/2}$ originates from the relation (\ref{tl}) 
between $\tau_2$ and $l$, after taking into account all implicit factors in the measure
over the moduli.

The transverse-channel annulus amplitude is determined by ${\cal T}_1$ of eq. (\ref{E3})
restricting the diagonal portion of the spectrum to the zero-momentum ($m=0$) sector.
Thus, only $V_8$ and $S_8$ with even windings and $I_8$ and $C_8$ with odd ones
are allowed in $\tilde{\cal A}$. We then associate to these terms the minimal number of
independent reflection coefficients, parametrized in terms of four integers
$(n_1,n_2,n_3,n_4)$, and write
\ba
\tilde{\cal A}_1 &=& \frac{2^{-{11/2}} R}{2}
\biggl( 
\left[ (n_1 + n_2 + n_3 + n_4 )^2 V_8 - ( n_1 + n_2 - n_3 - n_4 )^2 S_8 
\right] \tilde{Z}_{2n} \nonumber \\ &+& 
\left[ (n_1 - n_2 + n_3 - n_4 )^2 I_8 - ( n_1 - n_2 - n_3 + n_4 )^2 C_8 
\right] \tilde{Z}_{2n+ 1} \biggr) \ ,\label{E9}
\ea 
where from now on, for the sake of brevity, we shall often leave lattice 
sums implicit. This
parametrization solves the constraints of eq. (\ref{A}) and leads, via the $S_{(8)}$
transformation (\ref{E201}), to the direct-channel annulus amplitude 
\ba  
{\cal A}_1\! &=&\! \frac{n_1^2 + n_2^2 + n_3^2 + n_4^2}{2}  ( V_8 Z_{2m} - S_8 Z_{2(m+1/2)})
+ (n_1 n_2 + n_3 n_4 ) (V_8 Z_{2(m+1/2)} - S_8 Z_{2m} ) + \nonumber \\ 
\! &+&\! (n_1 n_3 + n_2 n_4
) (I_8 Z_{2m} - C_8 Z_{2(m+1/2)}) +  (n_1 n_4 + n_2 n_3 ) (I_8 Z_{2(m+1/2)} - V_8 Z_{2m}) \ ,
\label{E10}
\ea 
that describes an open spectrum with four types of Chan-Paton charges. As we shall
see below, additional reflection coefficients would correspond to the introduction of
Wilson lines.

Finally, the characters common to $\tilde{\cal  K}_1$ and $\tilde{\cal  A}_1$ determine 
the transverse M{\"o}bius amplitude
\ba
\tilde{\cal M}_1 = &-& \frac{R}{\sqrt{2}}
\biggl( (n_1 + n_2 + n_3 + n_4) \ {\hat V}_8 \
( {\tilde Z}_{4n} +  {\tilde Z}_{4n+2} )     
\nonumber \\ 
&-& (n_1 + n_2 - n_3 - n_4) \
{\hat S}_8 \ ({\tilde Z}_{4n} -  {\tilde Z}_{4 n+2} ) \biggr) \ , \label{E11}
\ea 
where, as in all subsequent M\"obius amplitudes, ${\tilde Z}_n$ (or $Z_m$) are
actually ``hatted'' Virasoro characters. 
The relative signs between the terms proportional to ${\tilde Z}_{4n}$ and to
${\tilde Z}_{4n+2}$, left undetermined by the geometric mean leading to 
$\tilde{\cal M}$, are
fixed  by the constraint (\ref{Mia}). Indeed, the naive choice of all equal
signs would lead to a M{\"o}bius amplitude incompatible with the particle 
interpretation of the open sector. This is
often the case in these constructions: 
eq. (\ref{Mia}) fixes the vacuum channel M{\"o}bius
amplitude $\tilde{\cal M}$ that,  in contrast to
$\tilde{\cal K}$ and  $\tilde{\cal A}$, is not restricted by positivity. The
direct-channel
M{\"o}bius amplitude is then obtained by the $P_{(8)}$ transformation (\ref{E201}), 
and is given by
\ba 
{\cal M}_1 = -\ \frac{ n_1 + n_2 + n_3 + n_4 }{2} \ {\hat V}_8 \ Z_{2m} +
\frac{ n_1 + n_2 - n_3 - n_4 }{2} \ {\hat S}_8 \ Z_{2(m+1/2)} \ . \label{E12}
\ea 
{}From eqs. (\ref{E10}) and (\ref{E12}), we see that the direct-channel M{\"o}bius
contribution is the proper (anti)symmetrization of the terms in the annulus amplitude
describing open strings with pairs of identical charges at their ends. Note that ${\cal
M}_1$ contains an additional factor 1/2 compared to $\tilde{\cal M}_1$. 
This rescaling is not a consequence of the
$P$-transformation, but is induced by the redefinition of the functional measure in
going from the transverse-channel variable $l \rightarrow l/2$.

The tadpole conditions are obtained from ${\tilde K}_1, {\tilde A}_1$ and  
${\tilde M}_1$, setting
to zero the total reflection coefficients for the massless modes, 
that for generic radii originate from $V_8$ and $S_8$, and read:
\ba
V_8 &:&\quad
{2^{9/2}\over 2}+{2^{-11/2}\over 2}(n_1 + n_2 + n_3 + n_4)^2-
{1\over{\sqrt 2}}(n_1 + n_2 + n_3 + n_4)=0\nonumber\\
S_8 &:&\quad
{2^{9/2}\over 2}+{2^{-11/2}\over 2}(n_1 + n_2 - n_3 - n_4)^2-
{1\over{\sqrt 2}}(n_1 + n_2 - n_3 - n_4)=0\ ,
\label{tadpoless}
\ea
so that
\ba  
n_1 + n_2 + n_3 + n_4 = 32 \ , \qquad n_1 + n_2 - n_3 - n_4 =32 \ . \label{E13}
\ea  
Thus, $n_3=n_4=0$ and $n_1+n_2=32$. It should be appreciated that all tachyons 
originating from the $I_8
Z_{2m}$ sector are then removed. The resulting open spectrum
\ba  
{\cal A}_1 &=& \frac{n_1^2 + n_2^2}{2} ( V_8 Z_{2m} - S_8 Z_{2(m + 1/2)} ) + n_1 n_2 (V_8
Z_{2(m + 1/2)} - S_8 Z_{2m} ) \ , \nonumber \\ 
{\cal M}_1 &=& - \frac{ n_1 + n_2 }{2} ( {\hat
V}_8 Z_{2m} - {\hat S}_8 Z_{2(m + 1/2)} ) \ , \label{E14}
\ea  
corresponds to a family of gauge groups $SO(n_1) \times SO(n_2)$, with
$n_1+n_2=32$. For interger momentum levels\footnote{According to our definitions
(\ref{tlA}), (\ref{hatt}) and eq. (\ref{Edef}), in the open spectrum $Z_{2m}$ describes
states of integer momenta with masses $m/R$, while $Z_{2(m+1/2)}$ describes states of
half-integer momenta.}, the spectrum consists of  vectors\footnote{In
9d the vector $V_8$ is a reducible representation, and already at the lowest level it
describes a vector
$A_{\mu}$ and a scalar $A_9$.} in the representations 
$({\bf n_1(n_1\!\!-\!\! 1)/2},{\bf 1})$ + $({\bf 1},{\bf n_2(n_2\!\! -\!\! 1)/2})$ and 
fermions in the representation $({\bf n_1},{\bf n_2})$. On
the other hand, for half-integer levels, the spectrum consists of fermions in the
$({\bf n_1(n_1\!\!-\!\! 1)/2},{\bf 1})$ + $({\bf 1},{\bf n_2(n_2\!\!-\!\!1)/2})$ 
and vectors in the $({\bf n_1},{\bf n_2})$. This partial
breaking results from Wilson lines \cite{BPS} 
in the original $SO(32)$ gauge group, a subject to which we now turn.

Our aim is now to compare this class of models with another obtained from the type-I
$SO(32)$ model breaking in part the gauge symmetry without affecting the supersymmetry
since, as we have anticipated, the model (\ref{E14}) originates from a discrete
deformation that induces the spontaneous breaking of supersymmetry. It is thus useful
to recall briefly how  Wilson lines can be introduced in open strings \cite{BPS},
considering the simplest example of the circle compactification of the type I
superstring to 9d. A Wilson line $W$ originating from a gauge field background 
$a=diag(a_1, -a_1 \cdots a_{16}, -a_{16})=(a_i)_{i=1 \cdots 32}$ in
the Cartan subalgebra of $SO(32)$, 
$W=diag(e^{2 \pi i a_1},e^{-2 \pi i a_1} \cdots e^{2 \pi i a_{16}}, e^{-2 \pi i
a_{16}})$, generically breaks $SO(32)$ to $U(1)^{16}$, while special values of the
parameters can break $SO(32)$ into products of even orthogonal subgroups\footnote{Odd
orthogonal subgroups can also be obtained, by discrete deformations with matrices 
$W$ in $O(32)$, with determinant equal to
$-1$.}. The resulting model is
\ba  
{\cal T}&=&|V_8-S_8|^2 Z_{m,n}\ , \qquad\qquad\qquad\ \   
{\cal K}={1 \over 2} (V_8-S_8) Z_m \ , \nonumber \\  
{\cal A}&=& {1 \over 2} (V_8-S_8)
\sum_{i,j=1}^{32} Z_{2(m+a_i+a_j)} \ , \qquad  
{\cal M}=-{1 \over 2} ({\hat V}_8-{\hat S}_8)
\sum_{i=1}^{32} Z_{2(m+2a_i)} 
\ . \label{E140}
\ea 
In the transverse channel the amplitudes take the elegant form
\ba  
{\tilde {\cal K}}&=&{2^{9/2} \over 2} \ R \ (V_8-S_8)  {\tilde Z}_{2n} \ ,
\nonumber \\ 
{\tilde {\cal A}}&=&{2^{-11/2} \over 2} \ R \ (V_8-S_8) 
\sum_n (Tr W^n)^2 {\tilde Z}_n \ , \nonumber \\  
{\tilde {\cal M}}&=&-{R \over {\sqrt 2}} ({\hat V}_8-{\hat S}_8)
\sum_n Tr W^{2n} {\tilde Z}_{2n} \ , \label{E141}
\ea 
where, for instance
\be
\sum_{i,j} e^{2 \pi i n (a_i+a_j)}= (Tr W^n)^2 \ . \label{E142}
\ee

We can thus break the gauge group to
$SO(n_1) \times SO(32-n_1)$ with $n_1$ even (or odd with a parity-like $W$) \cite{BPS},
and the resulting amplitudes
\ba 
{\cal K} &=& {1 \over 2} (V_8-S_8) Z_m \ , \nonumber \\  {\cal A} &=& (V_8-S_8)
\left (
\frac{n_1^2 + n_2^2}{2} Z_{2m} + n_1n_2 Z_{2(m+1/2)}
\right ) \ , \nonumber \\ {\cal M} &=& - \frac{ n_1 + n_2 }{2} ( V_8 - S_8) Z_{2m} \ ,
\label{E143} 
\ea 
show clearly that the model (\ref{E14}) is a discrete deformation (resulting in a
discrete mass shift, $m \to m + 1/2$, for the fermion spectrum) of the
supersymmetric model (\ref{E143}). This reflects the spontaneous character of the breaking,
that disappears in the limit $R
\to \infty$. In retrospect, this result is a natural extension of the boost (\ref{N11})
responsible for the Scherk-Schwarz breaking in the type-IIB string to the open sector.
Indeed, in the closed sector the boost may be seen as a left-right symmetric discrete
Wilson line built from the graviphoton. The discrete shifts (\ref{N12}) in the closed
fermion spectrum have thus counterparts in the open sector. Note that this model
describes also type I theory at finite temperature, upon the identification of the
compact coordinate with euclidean time, $T=1/(2\pi R)$.

One can also introduce continuous Wilson lines starting from 
the model (\ref{E14}) with
$n_2=0$ to induce various breakings of the gauge group. For instance, 16 pairs of different
parameters $(a_i,-a_i)$ lead to
\ba  {\cal A}_1 &=& \frac{1}{2}\sum_{i,j=1}^{32} 
( V_8 Z_{2(m+a_i+a_j)} - S_8 Z_{2(m + 1/2+a_i+a_j)} )
\ , \nonumber \\ {\cal M}_1 &=& - \frac{1}{2}\sum_{i=1}^{32} 
( V_8 Z_{2(m+2a_i)} - S_8 Z_{2(m + 1/2+2a_i)} ) \ ,
\label{E144}
\ea and the $SO(32)$ gauge group is correspondingly broken to $U(1)^{16}$. The models
(\ref{E14}) with even $n_2$ can be recovered as special cases if $n_2/2$ of the
$a_i$ equal $1/2$, $n_2/2$ equal $-1/2$ and the rest vanish. Alternatively, for arbitrary
$a_i$ this model corresponds to a Scherk-Schwarz discrete deformation of (\ref{E140}),
whereby the momenta of the fermionic modes are shifted by $1/2$ unit. As mentioned in
Section 3, this class of models is dual to corresponding  Scherk-Schwarz
compactifications of the $SO(32)$ heterotic string, whose gauge sector coincides with
(\ref{E144}) when restricted to zero windings.

For later use, we now show a simple instance of gauge symmetry breaking to
unitary groups, and thus a spectrum with 
oriented open strings. Starting again from $n_2=0$, this
corresponds to
$W = diag (e^{2 \pi i/4},e^{-2 \pi i/4}$ $\cdots$ $e^{2 \pi i/4}, e^{-2 \pi i/4})$, that
breaks
$SO(32)$ to $U(16)$. The resulting open sector is 
\ba  
{\cal A}_1 &=& n \bar{n} ( V_8 Z_{2m} - S_8 Z_{2(m+1/2)} ) + \frac{n^2 + \bar{n}^2}{2} 
(V_8 Z_{2(m+1/2)} - S_8 Z_{2m})  \ , \nonumber \\   {\cal M}_1 &=& - \frac{ n + \bar{n} }{2} (
V_8 Z_{2(m+1/2)} - S_8 Z_{2m} ) \ . 
\label{E18}
\ea Here $n (= \bar{n})$ label a pair of ``complex'' charges, corresponding to  the
fundamental representation of $U(n)$ and its conjugate \cite{BS}. 
The tadpole conditions then fix
$n=16$. Again,  there are no open-string tachyons. For integer momenta,
the spectrum consists of a vector in the adjoint and of a spinor in the antisymmetric 
${\bf n(n-1)/2}$ and its conjugate. For half-integer momenta, the spectrum consists of a
vector in the antisymmetric ${\bf n(n-1)/2}$ and its conjugate, and of a spinor in the
adjoint. This model, as well as special cases of eq. (\ref{E14}), 
were obtained previously (with double radius) in \cite{BD}, starting
from the orbifold  ${\cal T}_1^\prime$ of the IIB superstring in eq. (\ref{E5}).

\vskip 10pt
\begin{flushleft} {\large\bf 5.3. \ M-theory breaking model}
\end{flushleft}
\vskip 10pt
\noindent Starting from the torus amplitude ${\cal T}_2$ of eq. (\ref{E19}) with shifts
on the windings and following the same steps as before, one obtains
\be  
{\cal K}_2 = \frac{1}{2}  \ (V_8 - S_8) \ Z_{2m} + \frac{1}{2} \ (I_8 - C_8) \
Z_{2m+1}\ ,
\label{E20}
\ee and consequently
\be
\tilde{\cal K}_2 = \frac{2^{9/2}}{2} R ( V_8 
\tilde{Z}_{2n} -  S_8 \tilde{Z}_{2n+1} )\ . \label{E21}
\ee 
In contrast to the previous case, now the Klein bottle is affected by the
deformation.

The transverse annulus amplitude is determined from ${\cal T}_2$ of eq. (\ref{E19}),
restricting the diagonal portion of the spectrum to zero momentum $(m=0)$. There
are thus four independent reflection coefficients associated to $V_8$ and $S_8$ with 
even and with odd windings. As before, they can be parametrized
in terms of four integers $(n_1,n_2,n_3,n_4)$, so that
\ba
\tilde{\cal A}_2 &=& \frac{2^{-11/2}}{2} R
\biggl( 
\left[ (n_1 + n_2 + n_3 + n_4 )^2 V_8 - ( n_1 + n_2 - n_3 - n_4 )^2 S_8 
\right] {\tilde Z}_{2n} \nonumber \\ &+& 
\left[ (n_1 - n_2 + n_3 - n_4 )^2 V_8 - ( n_1 - n_2 - n_3 + n_4 )^2 S_8 
\right] {\tilde Z}_{2n+1} \biggr)\ , \label{E22}
\ea 
and the corresponding direct-channel annulus amplitude is
\ba 
{\cal A}_2 &=& \biggl( 
\frac{n_1^2 + n_2^2 + n_3^2 + n_4^2}{2} ( V_8 - S_8 ) 
+ (n_1 n_3 + n_2 n_4 ) ( I_8 - C_8 ) \biggr) Z_{2m} \nonumber \\ 
&+& \biggl( (n_1 n_2 + n_3 n_4 ) (V_8  - S_8 ) +  (n_1 n_4 +
n_2 n_3 ) (I_8 - C_8 )
\biggr) Z_{2(m+1/2)} \ . \label{E23}
\ea 
The characters common to ${\tilde K}_2$ and ${\tilde A}_2$ then determine  the
transverse-channel M{\"o}bius amplitude 
\ba
\tilde{\cal M}_2 &=& - \frac{R}{\sqrt{2}} \biggl( ( n_1 + n_2 + n_3 + n_4 ) \
{\hat V}_8 \ {\tilde Z}_{2n} - ( n_1 - n_2 - n_3 + n_4 )
\ {\hat S}_8  {\tilde Z}_{2n+1} \biggr) \ , \nonumber \\ 
{\cal M}_2 &=& - \frac{ n_1 + n_2 + n_3 + n_4 }{2} {\hat V}_8 Z_{2m} +
\frac{ n_1 - n_2 - n_3 + n_4 }{2} {\hat S}_8 (-1)^m Z_{2m} \ , \label{E24}
\ea  
where the implicit summation over $m$ now contains alternating signs in the last term.  

The tadpole conditions, obtained setting to zero the coefficients of the massless modes
originating from $V_8$ and $S_8$, give
\be n_1 + n_2 + n_3 + n_4 = 32 \ ,\qquad  n_1 + n_2 - n_3 - n_4 = 0 \ . \label{E25}
\ee Here, however, one needs further conditions to remove the tachyon from the $I_8 Z_m$
sector in $\cal A$, and a particularly interesting solution corresponds to $n_2=n_3=0$.
The result is the {\it unique} gauge group 
$SO(16) \times SO(16)$, and the open sector reads
\ba 
{\cal A}_2 &=&  
\frac{n_1^2 + n_4^2}{2} ( V_8 - S_8 ) Z_{2m} + n_1 n_4 (I_8 - C_8 ) Z_{2(m+1/2)} 
\ , \nonumber
\\ {\cal M}_2 &=& - \frac{ n_1 + n_4 }{2} {\hat V}_8 Z_{2m} +
\frac{ n_1 + n_4 }{2} {\hat S}_8 (-1)^m Z_{2m} \ . \label{E26}
\ea 
Notice that the massless open spectrum is actually supersymmetric, since
\ba {\cal A}_2 &=&  
\frac{n_1^2 + n_4^2}{2} ( V_8 - S_8 ) + {\rm massive} \ , \nonumber \\ 
{\cal M}_2 &=& -
\frac{ n_1 + n_4 }{2} ( {\hat V}_8 - {\hat S}_8 ) + {\rm massive} \ . 
\label{E27}
\ea 
Actually, the open spectrum is supersymmetric for even momenta, and 
in terms of $SO(8)$ representations describes a
vector and a spinor in the adjoint of $SO(16) \times SO(16)$. On the other hand, for 
odd momenta the vector is again in the adjoint, while the spinor is in the symmetric
representations $({\bf 135},{\bf 1})$$+$$2({\bf 1},{\bf 1})$$+$$({\bf 1},{\bf 135})$. 
Finally, for half-integer momenta,
there are scalars and spinors in the $({\bf 16},{\bf 16})$ representation.  

The duality arguments of Section 3 associate the closed sector of this type-I model,
after a T-duality, to a Scherk-Schwarz deformation affecting the momenta of the type-IIA
string. In the corresponding type-I$^\prime$ representation, however, the open strings
end on D8-branes perpendicular to the direction responsible for the breaking of
supersymmetry. Therefore, as we have just seen, all open string modes with even
windings, and in particular the massless ones, are unaffected. On the
other hand, both the changes in the open spectrum for the states with odd windings and
the presence of half-integer windings can be traced to corresponding modifications of
the closed spectrum due to the reversal of the GSO projection. This can be seen from eq.
(\ref{N15}), after interchanging $m$ and $n$. The soft nature of this breaking 
is less evident than in the previous example. In the next Section, however, we will
show that the one-loop contribution of the open string sector to the vacuum energy and
to the scalar masses is exponentially suppressed in the
decompactification limit, that in type I language corresponds to $R\to 0$.
Resorting again to the duality arguments of Section 3, it is clear that
this breaking corresponds to a non-perturbative phenomenon on the heterotic side and
realizes the Scherk-Schwarz deformation along the eleventh coordinate of M theory.

The tadpole equations (\ref{E25}) admit a second tachyon-free solution, $n_1=n_4=0$ and
$n_2=n_3=16$. It is then easy to see that the open-string sector is identical to that
of eq. (\ref{E26}), with $n_2$ and $n_3$ playing the role of $n_1$ and $n_4$, aside from
a crucial relative sign change between the ${\hat V}_8$ and ${\hat S}_8$ contributions
to ${\cal M}_2$. The resulting gauge group is again $SO(16) \times SO(16)$, but now
supersymmetry is broken already at the massless level, since the fermion
representations corresponding to even and odd values of momenta are interchanged. 
While this model is perturbatively consistent, it is
not clear to us what its dual M-theory interpretation is. Moreover, if one
requires, in the spirit of \cite{PW}, that all tadpoles that become massless
in the $R\to 0$ limit vanish, one is left with the unique choice $n_1=n_4=16$ 
and $n_2=n_3=0$, and the end result is the M-theory breaking model.
Furthermore, there are other
choices for the Klein-bottle projection for both models, that we do not
discuss here.

\section{One-loop scalar masses and the vacuum energy}
\vskip 10pt
\noindent
In the previous Section we have described a pair of type-I compactifications
obtained by Scherk-Schwarz deformations of momenta and windings. Since
in both models supersymmetry is broken, their spectra receive loop corrections.
In particular, we are interested in the loop corrections to the vacuum energy 
and to the masses of the states unaffected by supersymmetry breaking. 
The M-theory breaking model is particularly interesting in this respect, since
the residual global supersymmetry of its massless gauge sector is
expected to be broken only by gravitational interactions in the small radius 
limit.

In this Section we compute the one-loop mass corrections, in the limit 
of small supersymmetry breaking, for the
internal scalar components of ten-dimensional gauge fields, that remain massless to
lowest order. 
Gauge invariance simplifies this
task considerably: it makes these corrections universal, and allows one to deduce 
them from
the potential induced by generic Wilson lines. We thus confine our attention to
the scalar modes in the Cartan subalgebra, whose (constant) VEV's can be
identified  with open-string Wilson lines. In the
supersymmetric case, these define flat directions in the theory. In the broken case,
although they remain flat at the tree level,
quantum corrections are
expected to generate a potential, and we shall see that this is actually the case. 
Our
method then consists in computing the partition function (cosmological constant) 
in the
presence of Wilson lines, from which the masses can be obtained by differentiation 
at the
origin \footnote{Other nearby choices for the vacuum would generically supplement the
soft masses with supersymmetric contributions.}. 

The one-loop vacuum energy results from the contributions of the four surfaces of
vanishing Euler character:
\ba
\Lambda (W,R) &=& {1\over 2}\int_{\cal F} \frac{d^2 \tau}{\tau_2^{11/2}} \
\frac{{\cal T}(R)}{|\eta(\tau)|^{14}} + \int_{0}^{\infty} \ 
\frac{d \tau_2}{\tau_2^{11/2}} \ \frac{{\cal K}(R)}{\eta(2 i \tau_2 )^{7}} 
\nonumber \\ &+&  \int_{0}^{\infty} \ 
\frac{d t}{t^{11/2}}\ \frac{{\cal A}(W,R)}{\eta(i t/2 )^7} + 
\int_{0}^{\infty}  
\frac{d t}{t^{11/2}}\ \frac{{\cal M}(W,R)}{\hat{\eta}(i t/2 + 1/2)^7}
\ , \label{O1}
\ea
where $\cal F$ is the fundamental domain of the modular group for the world-sheet
torus, and for the Klein-bottle the modulus $t$ is the $\tau_2$ of the
previous Section. Furthermore,
$W$ denotes a generic Wilson line affecting the open sector and $R$ denotes the
radius of the circle. Referring to the two models of the previous Section, the limit
of small supersymmetry
breaking corresponds in the first case to $R \to \infty$, and in the second to
$R \to 0$. The torus contribution behaves as ${1/ R^9}$ in the first case and
as $R^9$ in the second~\footnote{For reductions to $d$ non-compact dimensions,
they would behave as ${1/ R^d}$ and as $R^d$, respectively.}. Therefore, in the
following we confine our attention to the remaining three contributions.
The scalar masses
may be obtained differentiating $\Lambda$ at the origin:
\be 
m_0^2 = R^2{\partial^2 \Lambda (W) \over \partial a \partial a}
\biggr|_{a=0} \ ,
\label{O2}
\ee 
where the factor $R^2$ is due to the normalization of the kinetic terms for the scalar
fields $a$ associated to the Wilson lines.

\vskip 10pt
\begin{flushleft} {\large\bf 6.1. \ Scherk-Schwarz breaking  model}
\end{flushleft}
\vskip 10pt
\noindent 
Without loss of generality, in this case we can start from the model with an
unbroken $SO(32)$ gauge group and introduce the Wilson line
$W= diag (e^{2 \pi i a_1},e^{-2 \pi i a_1},$ $\cdots$
$e^{2 \pi i a_{16}}, e^{-2\pi i a_{16}})$.
The Klein bottle is not affected \cite{BPS}, while
the corresponding annulus and M{\"o}bius direct-channel amplitudes 
are given in eq. (\ref{E144}).
In the transverse channel representation, one then finds
\ba 
{\tilde {\cal A}}_1 &=& {2^{-11/2} \over 2} R 
\sum_n \left[ (V_8-S_8) (TrW^{2n})^2 {\tilde Z}_{2n} + (I_8-C_8) (TrW^{2n+1})^2 
{\tilde Z}_{2n+1}
\right] \ , \nonumber \\ 
{\tilde {\cal M}}_1 &=& - {R \over \sqrt{2}} 
\sum_n \left[ {\hat V}_8-(-1)^n {\hat S}_8 \right] (TrW^{2n}) {\tilde Z}_{2n}\ . 
\label{O5}
\ea

It is convenient to express these results in terms of the cylinder length $l$, 
related to the direct-channel variables as in eqs. (\ref{tl}) and (\ref{tlA}).
Aside from the torus contribution, the cosmological constant then reads:
\be
\Lambda_{\rm extra} (W,R) = \int_0^{\infty} dl \ \left(
 \frac{{\tilde {\cal K}}_1(R)}{\eta(il)^7}+
\frac{{\tilde{\cal A}}_1(W,R)}{\eta(il)^7} + 
\frac{{\tilde{\cal M}}_1(W,R)}{\hat{\eta}{(il+1/2)^7}} \right) \ . 
\label{O9}
\ee 
In these models we are interested in the $R \rightarrow \infty$ limit, where
the dominant contribution comes from the $l \rightarrow 0$, infrared region of
integration. Using the ``abstruse identity'' $V_8=S_8$, the contribution of the
Klein bottle vanishes. On the other hand, the annulus and M{\"o}bius 
transverse-channel amplitudes take the form
\ba
\int_0^{\infty} dl \ \frac{{\tilde {\cal A}}_1}{\eta(il)^7} &=& {R \over 2^6 \sqrt {2}} 
\int_0^{\infty} dl {\theta_4^4 \over
\eta^{12}} (il) \sum_n (TrW^{2n+1})^2 e^{-{\pi l (2n+1)^2 R^2 \over 4 }}
\ \nonumber \\ 
\int_0^{\infty} dl \ \frac{{\tilde {\cal M}}_1}{\hat{\eta}(il+1/2)^7} &=& -{R \over \sqrt {2}} 
\int_0^{\infty} dl {\hat{\theta}_2^4 \over
\hat{\eta}^{12}} (il+{1\over 2}) \sum_n (TrW^{2(2n+1)}) e^{-{\pi l (2n+1)^2 R^2}} 
\ , \label{O10}
\ea 
and in the large-radius limit have the asymptotic behavior
\ba
\int_0^{\infty} dl \ \frac{{\tilde {\cal A}}_1}{\eta(il)^7}
&{{\Large \sim} \atop {R \to \infty}}& {1 \over 4 R^9 \sqrt {2}} \int_0^{\infty} dl \ l^4 \sum_n
(TrW^{2n+1})^2 e^{-{\pi l (2n+1)^2 \over 4}} 
\nonumber \\ 
\int_0^{\infty} dl \ \frac{{\tilde {\cal M}}_1}{\hat{\eta}(il+1/2)^7} 
&{{\Large \sim} \atop {R \to \infty}}& -{2^8 \over R^9 \sqrt {2}} 
\int_0^{\infty} dl \ l^4 \sum_n (TrW^{2(2n+1)}) e^{-{\pi l (2n+1)^2}} \ . 
\label{lmt}
\ea
In deriving the last expressions, we have rescaled the integration
variable according to $l \rightarrow l/R^2$, and we have extracted, after a modular
transformation, the dominant contribution of the string oscillators as $R \to \infty$. 

The integrals over $l$ are convergent and the result is
\be
\Lambda_{\rm extra} \simeq {2^{11} \ 3 \over \sqrt{2} \pi^5 R^9}\sum_n{1\over (2n+1)^{10}}
\left[ (TrW^{2n+1})^2- TrW^{2(2n+1)}\right]\ .
\label{res1l}
\ee
{}From eq. (\ref{O2}) the soft scalar masses then behave as $1/R^{7/2}$ 
in the large radius limit. As $W\to 1$, the factor tend to the usual field theory
factor, proportional to the number of particles circulating in the loop.
More generally, if the theory were compactified to $d$ dimensions,
one would obtain a vacuum energy $\Lambda$ 
scaling as ${1/R^d}$ and scalar masses scaling as $1/R^{(d-2)/2}$.
Therefore, in compactifications to four dimensions all soft masses are of the same 
order of magnitude
\be 
 m_0 \sim m_{1/2} \sim m_{3/2} \sim {1 \over R} \ , \label{O13}
\ee 
as expected from the effective field theory.

\vskip 10pt
\begin{flushleft} {\large\bf 6.2. \ M-theory breaking model}
\end{flushleft}
\vskip 10pt
\noindent 
Strictly speaking, in this case Wilson lines are not allowed.
This was pointed out in \cite{PW} for the supersymmetric model, 
where it was argued that the dilaton would otherwise develop strong
coupling singularities. We shall see shortly a different manifestation of the same
phenomenon: a divergent tadpole appears in 
the $R \to 0$ limit. This applies also to the Scherk-Schwarz deformed model.
With these qualifications, we turn on two sets of Wilson lines $W$ and $W^\prime$ in
the two factors of the gauge group.
The direct-channel annulus and M{\"o}bius amplitudes (\ref{E26}) then become
\ba 
{\cal A}_2 &=&{1 \over 2} (V_8-S_8) \sum_{i,j=1}^{16}( Z_{2(m+a_i+a_j)}+
Z_{2(m+a'_i+a'_j)})+ {1 \over 2} (I_8-C_8)\sum_{i,j=1}^{16} Z_{2(m+1/2+a_i+a'_j)}
\ , \nonumber \\ 
{\cal M}_2 &=&-{1 \over 2} \left [ {\hat V}_8-(-1)^m {\hat S}_8 \right]
\sum_{i=1}^{16} (Z_{2(m+2a_i)}+  Z_{2(m+2a'_i)}) \ , \label{014}
\ea 
while in the transverse channel the various amplitudes read
\ba 
{\tilde {\cal K}}_2&=&{2^{9/2} \over 2} R  (V_8 {\tilde
Z}_{2n}-S_8 {\tilde Z}_{2n+1}) \ , \nonumber \\ 
{\tilde {\cal A}}_2&=& {2^{-11/2} \over 2} R
\left[V_8 \sum_{n} (TrW^n+(-1)^n Tr{W^\prime}^n)^2 {\tilde Z}_n -S_8 (TrW^n-(-1)^n
Tr{W^\prime}^n)^2 {\tilde Z}_n \right] \ , \nonumber \\  
{\tilde {\cal M}}_2&=& -{R \over \sqrt{2}} 
\left[ {\hat V}_8 \sum_{n} (TrW^{2n}+ Tr{W^\prime}^{2n}) {\tilde Z}_{2n} -{\hat S}_8
(TrW^{2n+1}+Tr{W^\prime}^{2n+1}) {\tilde Z}_{2n+1} \right] 
 \ . \label{O15}
\ea 

In this model we are interested in the
$R \rightarrow 0$ limit. This is dominated by the $l \rightarrow \infty$, ultraviolet
region of integration, and is rather subtle, since as $R\to 0$ all winding states
flowing in the transverse channel contribute to massless tadpoles.
The tadpole conditions then read 
\be 
TrW^{2n}=Tr{W^\prime}^{2n}=16 \ , \qquad 
TrW^{2n+1}=Tr{W^\prime}^{2n+1}=16 \ , \label{0151}
\ee 
and demand that $W={W^\prime} = 1$, so that the $SO(16) \times SO(16)$ gauge group
is to remain unbroken. In fact, the same conditions apply to the supersymmetric model,
and provide a different manifestation of the phenomenon first noticed in
\cite{PW}. In the type-I$^\prime$ language, this limit
would naively correspond to weak coupling, but the dilaton actually develops
strong-coupling singularities at particular points on the circle \cite{PW}. 
From this perspective, the pathologically large vacuum values of the dilaton are 
clearly induced by
the new tadpoles resulting from the collapsed massive states. Therefore, 
a proper perturbative definition of the theory requires that
all the conditions (\ref{0151}) be satisfied.

After the rescaling $l \rightarrow l/R^2$, in the $R \rightarrow 0$ limit 
the transverse amplitudes read 
\ba 
\int_0^{\infty} dl \ \frac{{\tilde {\cal K}}_2}{\eta(il)^7} &=&  {2^{9/2} \over 2 R} 
\int_0^{\infty} dl \ \frac{\theta_2^4}{2 \eta^{12}}(il/R^2) \ 
(-1)^n e^{-{\pi l n^2 \over 4 }} \ , \label{O150} \\ 
\int_0^{\infty} dl \ \frac{{\tilde {\cal A}}_2}{\eta(il)^7} &=& {2^{-7/2} \over 2R} 
\int_0^{\infty} dl \ \frac{\theta_2^4}{2 \eta^{12}}(il/R^2) \ 
(-1)^n e^{-{\pi l n^2 \over 4 }} \ TrW^n \ Tr{W^\prime}^n \ , \nonumber \\  
\int_0^{\infty} dl \ \frac{{\tilde{\cal M}}_2}{\hat{\eta}(il+1/2)^7} &=&
 -{1 \over  \sqrt {2}R} 
\int_0^{\infty} dl \ \frac{\hat{\theta_2}^4}{2 \hat{\eta}^{12}} (il/R^2+ 1/2) \
(-1)^n e^{-{\pi l n^2 \over 4 }} (TrW^{n}+ Tr{W^\prime}^{n}) \nonumber
 \ , 
\ea 
where we have used the ``abstruse identity'' $V_8=S_8=\theta_2^4/2 \eta^4$.
Moreover, in this limit only the ground state contributes. Its degeneracy
can simply be accounted for letting 
$S_8/{\eta}^8, {\hat S}_8/{\hat \eta}^8 \rightarrow 8$, and thus
\ba 
\Lambda_{\rm extra} \ &{{\Large \sim} \atop {R \to 0}}& 
 {1 \over {2 \sqrt{2} R}}\int_0^{\infty} dl \ \sum_n Tr (W^n-1)
\ Tr ({W^\prime}^n-1) (-1)^n e^{-{\pi l n^2 \over 4 }} \nonumber\\
&=& {\sqrt{2} \over \pi R} \ \sum_n {(-1)^n \over n^2} Tr (W^n-1) \ Tr ({W^\prime}^n-1)\ .
\label{O16}
\ea
Therefore, the leading contribution to the potential $\Lambda_{\rm extra}$ is quartic
in the Wilson lines and the soft scalar masses vanish in this approximation, up to
exponentially suppressed corrections ${\cal O}(exp(-\pi/R))$. The latter come from the
massive oscillator modes of the string, that we have 
neglected. Furthermore, in this limit
the one-loop vacuum energy of the M-theory breaking model is not 
corrected to lowest order by $\Lambda_{\rm extra}$
and is given by the torus contribution, that behaves as $R^9$.
This result implies that the soft scalar masses are dominated by gravitational
contributions, that first arise from genus $3/2$ world-sheets, and by
field theory considerations are expected to give a contribution suppressed by powers
of the Planck mass.

\section{Supersymmetry breaking in six dimensions}

In Section 5 we have described two distinct 9d models with supersymmetry 
breaking induced by Scherk-Schwarz deformations affecting the momenta or the
windings. The generalization to lower dimensions involves, in general, the
simultaneous presence of D9 and D5 branes (modulo T-dualities). When only one
type of brane is present, the previous construction admits a straightforward
generalization that follows directly from the corresponding deformation of 
the torus amplitude. The purpose of this Section is to illustrate a
simple instance of supersymmetry breaking in six dimensions in the presence of
D9 and D5 branes. We should emphasize, however, that in this model the 
Scherk-Schwarz deformation is accompanied by a subtle change of chirality for
the twisted fermions already in the closed sector. This phenomenon, peculiar to
six dimensions, obscures somewhat the spontaneous nature of the breaking mechanism.

In six dimensions, supersymmetric open-string spectra have a rich structure,
due to the possible presence of D9 and D5 branes \cite{BS,GP,six}. By heterotic-type I
duality, nine branes correspond to perturbative gauge groups on the heterotic  
side, while five-branes correspond to zero-size heterotic instantons. As nine  
branes and five branes are T-dual of each other, the two basic ways of 
breaking supersymmetry discussed in Sections $4$ and $5$, with soft masses 
proportional to $1/R$ and to $R$, are expected to combine. 
In particular, the previous examples suggest that only one massless
sector (the $99$ one) feels supersymmetry breaking at tree
level, while the other massless sectors (the $59$ and the $55$ ones) 
can only be affected via radiative corrections.

In order to discuss the breaking of $(1,0)$ supersymmetry in  6d, we can start 
from
a six-dimensional IIB orbifold model and perform a Scherk-Schwarz 
deformation of the
closed string partition function in a way compatible with modular invariance, 
particle
interpretation and the orbifold projection. The original IIB massless 
spectrum has $(2,0)$
supersymmetry, and is uniquely fixed by anomaly cancellation: it consists of 
the gravitational multiplet, with 
$(g_{\mu\nu}, 2 \psi_{\mu L}, 5 B^+_{\mu\nu})$ and of 21
tensor multiplets, with $(B^-_{\mu\nu}, 2 \chi_R, 5 \phi )$, where 
$B^+$($B^-$) denote
tensors with (anti)self-dual field strengths and the chirality of the fermionic
fields is indicated by the $L$ and $R$ subscripts. 
The corresponding open descendants have
$(1,0)$ supersymmetry and, in addition to the gravitational multiplet 
$(g_{\mu\nu}, \psi_{\mu L}, B^+_{\mu\nu})$, contain in
general variable numbers of tensor multiplets $(B^-_{\mu\nu}, \chi_R, \phi )$,
vector multiplets $(A_{\mu}, \lambda_L)$ and hypermultiplets $(\psi_R,4\phi)$. 

To be concrete, let us consider in detail the
$T^4/Z_2$ type I orbifold of \cite{BS,GP} with all five-branes at the same
fixed point, with a single tensor multiplet, and with gauge group $U(16)_9
\times U(16)_5$. The open string sector then contains
vector multiplets and hypermultiplets in the representations 
$({\bf 120}+{\overline{\bf 120}},{\bf1})$, $({\bf1},{\bf 120}+{\overline{\bf 120}})$ 
of the gauge group from the $99$
and $55$ sectors, and one hypermultiplet in the representation 
$({\bf 16}, {\bf 16})$ from the $59$ sector.

{}Following \cite{BS}, let us introduce the convenient combinations of 
$SO(4)$ characters
\ba 
Q_O = V_4I_4-C_4C_4 \ , \qquad Q_V = I_4V_4-S_4S_4 \ , \nonumber\\ 
Q_S = I_4C_4-S_4I_4 \ , \qquad Q_C = V_4S_4-C_4V_4 \ , \label{S1} 
\ea
that allow to write the partition function in a compact form.
By definition, the first factor in (\ref{S1}) refers to the transverse 
coordinates of
spacetime, while the second refers to the internal (compact) ones. 
Thus, for instance,
$Q_O$ describes a Neveu-Schwarz vector and a corresponding Ramond $L$ spinor, 
$Q_V$  describes 4 scalars and a corresponding $R$ spinor, while 
the relative sign between
bosonic and fermionic contributions is dictated by spin statistics. In
terms of these characters, the torus amplitude of the $N=(2,0)$ supersymmetric
$T^4/Z_2$ orbifold reads
\ba 
{\cal T}&=&{1 \over 2} \Lambda^{(4,4)} |V_8-S_8|^2+{1 \over 2}  |Q_O-Q_V|^2
|I_4I_4-V_4V_4|^2_B
\nonumber \\ &+&{1 \over 2} \left\{ |Q_S+Q_C|^2 |Q_S+Q_C|^2_B +|Q_S-Q_C|^2
|Q_S-Q_C|^2_B \right\} \ , \label{S2}
\ea 
where $\Lambda^{(4,4)}$ denotes a $\Gamma_{(4,4)}$ Narain lattice \cite{N} for the
compact coordinates with a
vanishing antisymmetric tensor, and the subscript $B$ refers to the compact
bosonic modes, fermionized according to
\ba 4 {\eta^2 \over \theta_2^2} &=& {\theta_3^2 \theta_4^2 \over
\eta^4}=(I_4I_4-V_4V_4)_B
\ \ , \ \ 4 {\eta^2 \over \theta_4^2} = {\theta_2^2 \theta_3^2 \over \eta^4}=
(Q_S+Q_C)_B \ ,
\nonumber \\ 4 {\eta^2 \over \theta_3^2} &=& {\theta_2^2 \theta_4^2 \over \eta^4}=
(Q_S-Q_C)_B \ .
\label{S3}
\ea 

We can now break
supersymmetry deforming the partition function along a compact direction. 
To this end, let us specialize to a factorized 
$\Gamma_{(1,1)}\times\Gamma_{(3,3)}$ lattice and deform the first term of
eq. (\ref{S2}) according to
(\ref{N12}) with ${\bf e}=(0,0,0,1)$. This corresponds to the
operator $(-1)^F$ acting on the lattice states, and results from a
$2\pi$-rotation in a plane defined by one compact and one non-compact
coordinate. The corresponding world-sheet current anticommutes with the 
orbifold projection, as required for the consistency of the construction
\cite{KP,a}.  It is also convenient to introduce the additional characters
\ba 
Q'_O = V_4I_4-S_4S_4 \ , \qquad Q'_V = I_4V_4-C_4C_4 \ , \nonumber\\ Q'_S =
I_4S_4-C_4I_4
\ , \qquad Q'_C = V_4C_4-S_4V_4 \ , \label{S4} 
\ea 
so that the deformed partition function reads
\ba 
&{\cal T}& \!\!= \!\!{\Lambda^{(3,3)} \over 2}  
\left\{ E^{\prime}_0 ( |V_8|^2  \!+\! |S_8|^2  ) \!+\!  O^{\prime}_0 ( |I_8|^2  \!+\!
|C_8|^2 ) \!-\! E^{\prime}_{1/2} ( V_8 {\bar S}_8 \!+\!  S_8 {\bar V}_8 ) \!-\! 
O^{\prime}_{1/2} ( I_8 {\bar C}_8 \!+\! C_8 {\bar I}_8 ) \right\} \nonumber \\
\!\!&+&\!\!\!\!{1\over 4} (|Q_O\!-\!Q_V|^2+ |Q'_O\!-\!Q'_V|^2)
|I_4I_4\!-\!V_4V_4|^2_B \label{S5} \\
\!\!&+&\!\!\!\!{1\over 4}\left\{(|Q_S\!+\!Q_C|^2  \!+\! |Q'_S\!+\!Q'_C|^2) 
|Q_S\!+\!Q_C|^2_B\!+\! (|Q_S-Q_C|^2+ |Q'_S-Q'_C|^2) |Q_S-Q_C|^2_B \right\} \ . 
\nonumber
\ea 
Compared to the supersymmetric case (\ref{S2}), in the untwisted sector of the
deformed closed string (\ref{S5}) all fermion masses are shifted by
$1/(2R)$. On the other hand, as expected, in the twisted sector the fermion masses are
unchanged, while the chirality of half of the fermions is inverted. Eight of the twisted
massless
fermions are thus left handed, while the other eight are right handed. This
phenomenon will have a counterpart in the open sector. It originates from the
modifications introduced by the Scherk-Schwarz breaking at the origin of the lattice,
that eliminate all the corresponding fermionic modes. This affects the orbifold
projection in the untwisted sector and, by modular invariance, modifies the chirality of
the twisted states. This phenomenon is peculiar to
six-dimensional models, where closed string states can carry a net chirality, 
in contrast to the more familiar four-dimensional case. 

{}Following the same steps as in Section 5, the Klein bottle amplitudes in the
direct and transverse channels are
\ba 
{\cal K}\!\!\!\!&=&\!\!\!\!{1 \over 4} \left [ (V_8\!-\!S_8) (Z_m \Lambda^{(3)} 
\!+\! {\tilde Z}_{2n}
{\tilde
\Lambda}^{(3)})\!+\!(I_8\!-\!C_8){\tilde Z}_{2n+1}{\tilde
\Lambda}^{(3)}\!+\!(Q_S\!+\!Q_C\!+\!Q'_S\!+\!Q'_C) (Q_S\!+\!Q_C)_B
\right] ,
\nonumber \\ {\tilde {\cal K}}\!\!\!\!&=&\!\!\!\!{2^5 \over 4} 
\left [ v_4 (V_8-S_8)  {\tilde Z}_{2n}{\tilde \Lambda}_e^{(3)}+{1 \over v_4}
(V_8Z_{2m}-S_8Z_{2m+1})
\Lambda_e^{(3)} \right]
\nonumber \\ \!\!&+&\!\!{2^5 \over 4} (Q_O-Q_V+Q'_O-Q'_V) (I_4I_4-V_4V_4)_B  
\ , \label{S6}
\ea 
where $v_4$ is the volume of the compact space, $\Lambda^{(3)}$ contains only momenta
and ${\tilde\Lambda}^{(3)}$ contains only windings.  Moreover, in $\Lambda_e^{(3)}$ and
${\tilde \Lambda}_e^{(3)}$ the lattice sums are restricted to even values of momenta and
windings, respectively. 

It is instructive to exhibit in $\tilde{\cal K}$ the general
structure of eq. (\ref{Ktilde}). While evident for the contributions of all lattice
points away from the origin, this is not apparent for the remaining states.
To this end, let us recall the decomposition of level-one $SO(8)$
characters into $SO(4)$ ones~,
\ba
&I_8 = I_4 I_4 + V_4 V_4 \ , \qquad &C_8 = S_4 C_4 + C_4 S_4 \quad ,
\nonumber \\
&V_8 = V_4 I_4 + I_4 V_4 \ , \qquad &S_8 = S_4 S_4 + C_4 C_4 \quad ,
\label{break}
\ea
that may be simply induced from elementary group embeddings. Leaving aside all terms away
from the origin of the lattice, $\tilde{\cal K}$ reduces to
\ba 
{\tilde {\cal K}}_0 \!&=&\!{2^5 \over 4} \left[ v_4 
(V_4I_4+I_4V_4-S_4S_4-C_4C_4) +
{1 \over v_4} (V_4I_4+I_4V_4) \right] (I_4I_4+V_4V_4)_B \nonumber \\ \!&+&\!{2^5 \over
4} 2 (V_4I_4-I_4V_4) (I_4I_4-V_4V_4)_B \ , \label{S9}
\ea 
where in the first term $(I_4I_4+V_4V_4)_B$ denotes the contribution from the origin of
the Narain lattice. This expression can then be reassembled in the form
\ba 
{\tilde {\cal K}}_0 \!&=&\!{2^5 \over 4} ( {\sqrt v_4} +{1 \over \sqrt
v_4})^2 
\left[ V_4I_4(I_4I_4)_B+I_4V_4(V_4V_4)_B \right]  \nonumber \\ \!&+&\!{2^5 \over 4} (
{\sqrt v_4} -{1
\over \sqrt v_4})^2 
\left[ V_4I_4(V_4V_4)_B+I_4V_4(I_4I_4)_B \right] \nonumber \\ \!&-&\!{2^5 \over 4} 
({\sqrt v_4})^2 (S_4S_4+C_4C_4) (I_4I_4+V_4V_4)_B \ , \label{S10} 
\ea 
where all coefficients are indeed perfect squares.

As in Section 5, the transverse-channel annulus amplitude results from 
contributions of the different sectors of the closed string, each weighted by
a corresponding (squared) reflection coefficient.  The non-vanishing 
coefficients
pertain to the twisted states, as well as to the untwisted ones belonging to
two sublattices of the original Narain lattice.  In particular, terms with zero 
momentum correspond to Neumann boundary conditions, while terms with 
zero winding correspond to Dirichlet ones. These will shortly pair with corresponding
terms in $\tilde{\cal K}$ proportional to the internal volume $v_4$ and to its inverse.
Furthermore, the reflection coefficients are parametrized, as usual, in terms of 
some integers
that we shall shortly relate to Chan-Paton multiplicities. For untwisted
closed-string states, these comprise
$n_N$, the dimensionality of the Chan-Paton space for the 
(Neumann) 9-brane charges and $( n_{D_1},n_{D_2} )$, the corresponding 
dimensionalities for two types of (Dirichlet) 5-brane charges. In addition,
the reflection coefficients for twisted closed-string states are parametrized in
terms of the integers $R_N$, $R_{D_1}$ and $R_{D_2}$ \cite{PS}. One thus finds:
\ba 
2^7 {\tilde {\cal A}}\!&=&\! v_4 n_N^2 \biggl( (V_8-S_8){\tilde Z}_{2n} +(I_8-C_8)
{\tilde Z}_{2n+1}
\biggr) {\tilde \Lambda}^{(3)} \nonumber \\  \!&+&\! {1 \over v_4} \biggl( 
(n_{D_1}+n_{D_2})^2 (V_8 Z_{2m}-S_8 Z_{2m+1})+(n_{D_1}-n_{D_2})^2 (V_8  Z_{2m+1}-S_8
Z_{2m}) \biggr)
\Lambda^{(3)}  \nonumber \\ \!&+&\! \biggl( 2n_Nn_{D_1}(Q'_O-Q'_V)+ 2n_Nn_{D_2}(Q_O-Q_V)
\biggr) (I_4I_4-V_4V_4)_B
\nonumber \\ \!&+&\! \biggl( 2 R_N^2 (Q_S+Q_C+Q'_S+Q'_C)+4R_{D_2}^2 (Q_S+Q_C)+4R_{D_1}^2
(Q'_S+Q'_C)
\biggr) (Q_S+Q_C)_B \nonumber \\ \!&+&\! \biggl( 2R_NR_{D_2}(Q_S-Q_C)+
2R_NR_{D_1}(Q'_S-Q'_C)
\biggr) (Q_S-Q_C)_B  \quad . \label{S8}
\ea 

Just like $\tilde{\cal K}$, the transverse-channel annulus amplitude $\tilde{\cal A}$
satisfies a rather stringent consistency condition: the reflection
coefficients of the various sectors of the closed spectrum
are to be perfect squares, as in eq. (\ref{Atilde}). 
Again, this property is not apparent for the contributions arising
from the origin of the Narain lattice and from twisted
sectors.  It is therefore instructive to display these terms, that we denote
collectively by ${\tilde{\cal A}}_0$, in the properly reassembled form:
\ba 
2^7 {\tilde {\cal A}}_0 \!\!&=&\!\! ( {\sqrt v_4}n_N \!+\!{n_{D_1}+n_{D_2}
\over \sqrt v_4})^2 
\left[ V_4I_4(I_4I_4)_B\!+\!I_4V_4(V_4V_4)_B \right]   \nonumber \\ \!&+&\! (
{\sqrt v_4}n_N -{n_{D_1}+n_{D_2} \over \sqrt v_4})^2 
\left[ V_4I_4(V_4V_4)_B+I_4V_4(I_4I_4)_B \right] \ \ \ \ \ \ \nonumber \\ \!&-&\! (
{\sqrt v_4} n_N +{n_{D_1}-n_{D_2} \over \sqrt v_4})^2 
\left[ S_4S_4 (I_4I_4)_B+C_4C_4 (V_4V_4)_B \right]  \ \ \ \  \nonumber \\ \!&-&\! (
{\sqrt v_4}n_N -{n_{D_1}-n_{D_2} \over \sqrt v_4})^2 
\left[ S_4S_4 (V_4V_4)_B+C_4C_4 (I_4I_4)_B \right] \ \ \ \ \label{S11} \\ 
\!&+&\! \biggl(
{(R_N+4R_{D_2})^2 \over 4}+{7  R_N^2 \over 4} \biggr) (Q_S Q_{SB}+Q_CQ_{CB}) \nonumber
\\
\!&+&\!
\biggl(  {(R_N-4R_{D_2})^2 \over 4}+ {7 R_N^2 \over 4} \biggr) (Q_S Q_{CB}+Q_CQ_{SB})
\nonumber
\\ \!&+&\!
\biggl(  {(R_N+4R_{D_1})^2 \over 4} +{7  R_N^2 \over 4} \biggr) (Q'_S
Q_{SB}+Q'_CQ_{CB}) \nonumber \\\!&+&\! \biggl(  {(R_N-4R_{D_1})^2 \over 4}+ {7 R_N^2
\over 4}
\biggr) (Q'_S Q_{CB}+Q'_CQ_{SB})
 \ . \nonumber  
\ea
The last four lines are particularly interesting, since they describe the reflections
of the twisted
sectors of the closed string. As is well known, these are confined to the 16 fixed
points of the orbifold, and thus eq. (\ref{S11}) contains a rather detailed information
on the geometry of the D5 branes.  There is a little complication, however, since
we are actually describing the full twisted sector in terms of two pairs of characters
$(Q_S,Q_C)$ and $(Q^\prime_S,Q^\prime_C)$. While these suffice to identify
two types of fixed points, they clearly cannot distinguish any two fixed points of the
same type. As a result, in this case the reflection coefficients are
actually sums of squares. Still, their interpretation is 
quite transparent. The terms related to the pair $(Q_S,Q_C)$ imply that all
D5 branes of one type are concentrated
in one of eight available fixed points, and
the terms related to the pair 
$(Q^\prime_S,Q^\prime_C)$ imply a similar distribution for the second type of
D5 branes among the remaining eight fixed points.  To reiterate, out of the 16
fixed points of the orbifold 14 are empty, and the corresponding sectors of the
closed string sense only the ubiquitous nine-branes. One of the remaining two
fixed points accommodates all the D5 branes of one type, while the other 
accommodates all those of the second type. Actually, as
we shall see shortly, the tadpole conditions imply that 
the model includes a total of $16$ D5 branes, as does the standard
version with unbroken supersymmetry, 
so that eight of them are actually located at each of these two fixed points.

A modular $S$ transformation determined by eq. (\ref{E201}) yields the
direct-channel annulus amplitude
\ba 
{\cal A}\!\!\!\!&=&\!\!\!\! 
{n_N^2\over 4} (V_8Z_{2m}\!-\!S_8Z_{2(m+1/2)}\!) \Lambda^{(3)} \!+\! 
\!\!{1\over 4}\!\!\biggl( \!\!
(n_{D_1}^2\!+\!n_{D_2}^2)(V_8\!-\!S_8){\tilde Z}_{2n}\!+\! 2 n_{D_1}n_{D_2} 
(\! I_8\!-\!C_8)
{\tilde Z}_{2(n+1/2)} \!\!\biggr)\! {\tilde \Lambda}^{(3)}_0 \nonumber \\ \!\!&+&\!\! 
{1\over 4}\biggl(
2n_Nn_{D_1}(Q'_S+Q'_C)+ 2n_Nn_{D_2}(Q_S+Q_C) \biggr) ({Q_S+Q_C \over 4})_B 
\nonumber \\ 
\!\!\!\!&+&\!\!\!\! {1\over 4}\biggl( {1 \over 2} R_N^2 (Q_O-Q_V+Q'_O-Q'_V)+R_{D_2}^2 
(Q_O-Q_V)+R_{D_1}^2 
(Q'_O-Q'_V) \biggr) (I_4I_4\!-\! V_4V_4)_B \nonumber \\ \!\!&+&\!\! {1\over 4}\biggl(
2R_NR_{D_2}(Q_S-Q_C)+ 2R_NR_{D_1}(Q'_S-Q'_C)
\biggr) ({{Q_S-Q_C} \over 4})_B \ . 
\label{S7}
\ea 
As in \cite{PS},
the $R$ terms describe the combined action of the orbifold
involution on the open-string sectors and on the corresponding
Chan-Paton charges. In our case, the orbifold involution
consists of a pair of $\pi$-rotations in the $(6,7)$ and
$(8,9)$ planes that split the contributions to eq. (\ref{break}),
inverting the signs of the terms involving the 
internal $V_4$ and $S_4$.  The $R$ terms effect a corresponding 
splitting in the Chan-Paton charge space.  There are actually two
distinct options in this case, that correspond to ``real'' and
``complex'' charges or, equivalently, to orthogonal or 
symplectic and to unitary groups, respectively.  In the former case,
$N = n_+ + n_-$, 
and
\be
R = n_+ - n_- \qquad ,\label{real}
\ee
where $n_+$ and $n_-$ are identified with the charge multiplicities in
the unbroken gauge groups. In the latter case \cite{BS}, $N = n+ \bar{n}$, and 
\be
R = i(n - \bar{n}) \qquad ,\label{complex}
\ee
where $n$ and $\bar{n}$ are identified with the (identical) charge
multiplicities of the fundamental and conjugate representations of a
unitary group.  Indeed, one can simply see that 
in this case the positivity constraints on the
transverse-channel annulus amplitude require that $n=\bar{n}$.

The geometric means of the reflection coefficients for the
sectors common to ${\tilde {\cal K}}$ and ${\tilde {\cal A}}$ 
now determine the transverse-channel M{\"o}bius amplitude:
\ba 
-2 {\tilde {\cal M}}\!\!\!&=&\!\!\! n_N v_4 \biggl( {\hat V}_8 
({\tilde Z}_{4n}\!+\! {\tilde
Z}_{4n+2})\!-\!{\hat S}_8 ({\tilde Z}_{4n}\!-\!{\tilde Z}_{4n+2}) \biggr) {\tilde
\Lambda}_e^{(3)}\!+\!\frac{(n_{D_1}\!+\!n_{D_2})}{v_4} 
({\hat V}_8Z_{2m}\!-\!{\hat S}_8
Z_{2m+1})\Lambda_e^{(3)} 
\nonumber \\ 
\!\!\!&+&\!\!\! \biggl( n_N (V_4I_4\!-\!I_4V_4)\!+\!n_{D_1}({\hat Q'}_O\!-\!{\hat
Q'}_V) +n_{D_2}({\hat Q}_O\!-\!{\hat Q}_V) \biggr) 
({\hat I}_4 {\hat I}_4 \!-\!{\hat V}_4
{\hat V}_4)_B  \quad  . \label{S12}
\ea 
The reader can easily verify that the terms at the origin of the lattice,  
${\tilde {\cal M}}_0$, take the form of eq. (\ref{Mtilde}). Indeed,
starting from eq. (\ref{S12}) one gets
\ba 
-2 {\tilde {\cal M}}_0 \!&=&\! \left( n_N v_4 ( {\hat V}_8 -{\hat S}_8) + {1 \over
v_4} (n_{D_1}+n_{D_2}) {\hat V}_8 \right)
({\hat I}_4 {\hat I}_4 \!+\!{\hat V}_4 {\hat V}_4)_B
\nonumber \\ \!\!&+&\!\! \biggl( n_N
({\hat V}_4{\hat I}_4\!-\!{\hat I}_4{\hat V}_4)\!+\!n_{D_1}({\hat Q'}_O\!-
\!{\hat Q'}_V) \!+\!n_{D_2}({\hat Q}_O\!-\!{\hat Q}_V)
\biggr) ({\hat I}_4 {\hat I}_4 \!-\!{\hat V}_4 {\hat V}_4)_B  \
\label{S13}   
\ea 
that, using the decompositions (\ref{break}), becomes:
\ba
-2 {\tilde {\cal M}}_0 \!&=&\! ({\sqrt v_4} +{1 \over \sqrt v_4})
({\sqrt v_4}n_N \!+\!{n_{D_1}+n_{D_2}\over \sqrt v_4})
\left[ {\hat V}_4{\hat I}_4({\hat I}_4{\hat I}_4)_B+
{\hat I}_4{\hat V}_4({\hat V}_4{\hat V}_4)_B \right] 
\nonumber \\ \!&+&\! 
({\sqrt v_4} -{1\over \sqrt v_4}) 
({\sqrt v_4}n_N \!-\!{n_{D_1}+n_{D_2}\over \sqrt v_4})
\left[ {\hat V}_4{\hat I}_4({\hat V}_4{\hat V}_4)_B+
{\hat I}_4{\hat V}_4({\hat I}_4{\hat I}_4)_B \right] 
\nonumber \\ \!&-&\!
{\sqrt v_4}({\sqrt v_4} n_N +{n_{D_1}-n_{D_2} \over \sqrt v_4}) 
\left[ {\hat S}_4{\hat S}_4 ({\hat I}_4{\hat I}_4)_B+
{\hat C}_4{\hat C}_4 ({\hat V}_4{\hat V}_4)_B \right]  \ \ \ \  
\nonumber \\ \!&-&\! 
{\sqrt v_4}({\sqrt v_4}n_N -{n_{D_1}-n_{D_2} \over \sqrt v_4}) 
\left[ {\hat S}_4{\hat S}_4 ({\hat V}_4{\hat V}_4)_B+
{\hat C}_4{\hat C}_4 ({\hat I}_4{\hat I}_4)_B \right] \ .
\ \ \ \ \label{S101} 
\ea
The $P_{(4)}$ matrix of eq. (\ref{E201}) then determines the
direct channel M{\"o}bius amplitude~:
\ba 
{\cal M} \!&=&\! - {n_N\over 4} ({\hat V}_8  Z_{2m}\!-\!{\hat S}_8 Z_{2(m+1/2)})
 \Lambda^{(3)}- {n_{D_1}+n_{D_2}\over 4} 
 ({\hat V}_8 {\tilde Z}_{2n}\!-\!{\hat S}_8 (-1)^n  {\tilde
Z}_{2n}) {\tilde
\Lambda}^{(3)}_0  \nonumber \\ 
\!\!&+&\!\! {1\over 4}\biggl( n_N (V_4I_4\!-\!I_4V_4)+n_{D_1}({\hat
Q'}_O\!-\!{\hat Q'}_V) +n_{D_2}({\hat Q}_O\!-\!{\hat Q}_V) \biggr) ({\hat I}_4 
{\hat I}_4\!-\!{\hat V}_4 {\hat V}_4)_B  \ .
\label{S14}
\ea 

The tadpole equations related to twisted states demand that
\be R_N \ = \ R_{D_1} \ = \ R_{D_2} \ = \ 0 \ , \label{S15}
\ee  
since the corresponding characters appear only in the annulus. Three
more tadpole conditions originate from the massless
contributions of untwisted states, that add up to
\ba
& &\frac{2^5}{4} \left[ \sqrt{v_4}\left( 1 - \frac{n_N}{32} \right)
+ \frac{1}{\sqrt{v_4}}\left(1 -
  \frac{n_{D1}+n_{D2}}{32}\right)\right]^2
V_4 I_4 (I_4 I_4 )_B  \nonumber \\ &+&\frac{2^5}{4} 
\left[ \sqrt{v_4}\left( 1 - \frac{n_N}{32} \right)
+-\frac{1}{\sqrt{v_4}}\left(1 -
  \frac{n_{D1}+n_{D2}}{32}\right)\right]^2
I_4 V_4 (I_4 I_4 )_B \nonumber \\
&-&\frac{2^5}{4} \left[\sqrt{v_4} \left(1 - \frac{n_N}{32} \right) + 
\frac{n_{D1}-n_{D2}}{\sqrt{v_4}} \right]^2 S_4 S_4 (I_4 I_4)_B \nonumber \\
&-&\frac{2^5}{4} \left[\sqrt{v_4} \left(1 - \frac{n_N}{32} \right) + 
\frac{n_{D2}-n_{D1}}{\sqrt{v_4}} \right]^2 S_4 S_4 (I_4 I_4)_B \quad .
\label{tadpoleterms}
\ea
One thus finds
\be n_N = 32 \ , \qquad n_{D_1} =16 \,  \qquad  n_{D_2} = 16 \ . \label{S17}
\ee  
In view of (\ref{S15}), the $Z_2$ splittings of the Chan-Paton charge spaces
can describe ``complex'' charges associated to unitary groups, 
as in eq. (\ref{complex}). Thus, letting
\ba 
& &n_N = n + {\bar n} \ ,\qquad\qquad\quad 
R_N = i(n - {\bar n}) \ ,\nonumber \\ & &n_{D_1} = m_1 +
{\bar m}_1
\ ,\qquad\quad\  n_{D_2} = m_2 + {\bar m}_2 \ ,\nonumber \\  
& &R_{D_1} = i(m_1 - {\bar m}_1) \ , \qquad R_{D_2} =
i(m_2 - {\bar m}_2) \ , \label{S18}
\ea 
one finds $n=16$, $m_1=8$, $m_2=8$, and the resulting gauge group is $U(8)_5
\times U(8)^\prime_5 \times U(16)_9$, where the two $U(8)$ factors
originate from the D5 branes.

The massless open-string spectrum can be read from the
relevant parts of eqs. (\ref{S7}) and (\ref{S14}), that are
\ba
{\cal A} &=& n {\bar n} \ V_4 I_4 + m_1 {\bar m_1}(V_4 I_4 - S_4 S_4)
+ m_2 {\bar m_2} (V_4 I_4 - C_4 C_4) \nonumber \\
&+& \frac{n^2+{\bar n}^2}{2} \ I_4 V_4
 + \frac{m_1^2+{\bar m_1}^2}{2} \ (I_4V_4 -
 C_4 C_4)
 + \frac{m_2^2+{\bar m_2}^2}{2} \ (I_4V_4 -
 S_4 S_4) \nonumber \\ 
&+& (n \bar{m}_1+ \bar{n} m_1 ) Q_S^\prime + (n \bar{m}_2 + \bar{n} m_2 )
Q_S\\
{\cal M} &=& -  \frac{n+{\bar n}}{2} \ \hat{I}_4 \hat{V}_4
- \frac{m_1+{\bar m_1}}{2} \ (\hat{I}_4 \hat{V}_4 -
 \hat{C}_4 \hat{C}_4)
 - \frac{m_2+{\bar m_2}}{2} \ (\hat{I}_4 \hat{V}_4 -
 \hat{S}_4 \hat{S}_4)
\ea
Thus, in the $99$ (NN) sector, aside from the gauge bosons of
$U(16)_9$, there are quartets of scalars in the ${\bf
120}+{\bf \overline{120}}$ representations, while the corresponding 
fermionic modes, massless in the supersymmetric $U(16) \times U(16)$ model, 
are now massive as a result of supersymmetry breaking. In
the $55$ (DD) sector, aside from the gauge bosons of  
$U(8)_5 \times U(8)'_5$ and the corresponding adjoint fermions, 
$({\bf 64},{\bf 1})_R$ and $({\bf 1},{\bf 64})_L$,
there are quartets of
scalars and corresponding fermions in the representations 
$({\bf 28},{\bf 1})_L$, $({\bf \overline{28}},{\bf 1})_L$, 
$({\bf 1}, {\bf 28})_R$ and $({\bf 1}, {\bf \overline{28}})_R$. 
Finally, the states of the $59$ (ND) sector are in mixed representations 
of the full gauge group $U(8)_5 \times U(8)'_5 \times U(16)_9$.  
The corresponding massless spectrum consists of pairs of scalars and
corresponding (half)fermions\footnote{This peculiar feature reflects 
the pesudoreality of six-dimensional Weyl spinors.} in the representations $({\bf
{\bar 8}},{\bf 1},{\bf 16})_L$ and $({\bf 1},{\bf 8},
{\bf \overline{16}})_R$.

Notice that, in analogy with the twisted sector of the closed string, 
the 55 and 59 sectors of the open spectrum contain even numbers of
fermionic and bosonic modes at the massless level. 
This is just the phenomenon described in Section 5,
since the 5-brane world-volume is
orthogonal to the coordinate that breaks supersymmetry. 
As we already emphasized, however, there is a change of chirality for all fermions
charged under the $U(8)'_5$ gauge group, that reflects the corresponding 
phenomenon in the closed-string sector. 

Finally, using the results of \cite{AGW}, one can compute the anomaly polynomial, that clearly contains no
irreducible $R^4$ term, since there is no net number of chiral fermions or tensors.  
Moreover, the tadpole conditions
eliminate the irreducible $F^4$ terms, and the residual anomaly polynomial,
\be
A = \frac{1}{4} ({\rm tr}F_5^2 - {\rm tr}F_{5^\prime}^2 ) ({\rm tr}F_9^2 + 
\frac{1}{2} {\rm tr} R^2) \quad ,
\label{residual}
\ee
is rather similar to the corresponding one of the supersymmetric 
$U(16) \times U(16)$ model.

\section{Conclusions}

In this paper we have studied supersymmetry breaking by compactification 
in open descendants of
the type IIB closed string theory, using a generalization to
superstrings of the Scherk-Schwarz mechanism. 
We have exhibited two
basic ways of realizing the breaking, that for the type IIB 
parent models originate from shifts of momenta and windings, respectively. 
The corresponding type-I descendants exhibit vastly different properties:
in the former case the result is a rather direct extension to
the open sector of the usual mechanism, while in the latter the massless open modes
are not affected.
Moreover, at tree level the first mechanism gives supersymmetry breaking masses
of order $1/R$ both in the closed and in the open sector. This is the case, since
the direction used
to implement the Scherk-Schwarz deformation 
is parallel to the worldvolume of the open-string branes. The
corresponding models can be viewed as discrete deformations of
supersymmetric ones containing Wilson lines, and are
perturbatively dual to heterotic models with appropriate
Scherk-Schwarz breaking. This correspondence was anticipated by duality
arguments in Section 3.

In the second, qualitatively different mechanism, the coordinate used to
implement the breaking is orthogonal to the
worldvolume of the open-string branes, and in the closed sector the
resulting soft masses are of order $R/\alpha^\prime$.
The duality arguments of Section 3 associate this breaking 
to the eleventh dimension of M theory, and suggest that the
massless open spectrum should not be affected at tree
level. This is indeed confirmed by the explicit construction of Section 5.
On the other hand, the massive open spectrum is affected by 
the breaking, that at one-loop is communicated to the massless states.
The corresponding corrections, however, are exponentially suppressed in
the radius as $e^{-1/R}$. Therefore, the
primary source of breaking in the open string massless spectrum are the
gravitational interactions, expected to 
generate soft masses $\simlt{\cal O}(m_{3/2}^2/M_P)$. 

Type I models compactified to six and lower dimensions generically
contain 9-branes and 5-branes.  Since a generic compact coordinate is 
only parallel to the worldvolume of some of the
branes, both mechanisms are expected to play a role. As we have seen,
supersymmetry breaking along this coordinate affects at tree-level both the closed
and the open spectrum of the branes parallel to it. On the other hand,
the massless open spectra of branes orthogonal to
this coordinate and from mixed sectors 
feel the breaking only through radiative corrections. When 
the corrections are purely of gravitational origin,
the scale of supersymmetry
breaking in the ``orthogonal" branes is highly suppressed. Thus, one can attain
phenomenologically
interesting models with intermediate values for the compactification scale, 
of order $10^{12}-10^{14}$ GeV. Furthermore, the resulting scenario 
is compatible with gauge-coupling
unification at the grand unified scale, identified also with the
string scale  $M_I \sim 10^{16} GeV$. We emphasize that this
mechanism is nonperturbative from the heterotic string
point of view. 
It is certainly important to extend this construction to chiral
four dimensional type I models \cite{ABPSS}, and to study its implications.
\\[5mm]
\noindent{\bf Acknowledgements} 

It is a pleasure to acknowledge stimulating discussions with
C. Angelantonj, E. Kiritsis, C. Kounnas and H. Partouche.
E.D. and A.S. would like to thank the Centre de Physique Th\'eorique of the Ecole
Polytechnique and the Laboratoire de Physique Th\'eorique of the Ecole 
Normale Sup\'erieure. A.S. would like to thank the Theory Division of CERN for
the kind hospitality while this work was being completed, and 
M. Bianchi, G. Pradisi and Ya.S. Stanev, with whom the factorized form of
the vacuum amplitudes was exhibited long ago in unpublished work.



\begin{thebibliography}{99}

\bibitem{a} I. Antoniadis, \PLB{246}{90}{377};\\ Proc.
PASCOS-91 Symposium, Boston (World Scientific, Singapore, 1991), p.~718;\\
K. Benakli, \PLB{386}{96}{106}.

\bibitem{AMQ} I. Antoniadis, C. Munoz and M. Quiros, \NPB{397}{93}{515};\\
I. Antoniadis and K. Benakli, \PLB{326}{94}{69};\\ 
I. Antoniadis, K. Benakli and M. Quir\'os, \PLB{331}{94}{313};\\
I. Antoniadis, S. Dimopoulos and G. Dvali, \NPB{516}{98}{70}.

\bibitem{SS} J. Scherk and J.H. Schwarz, \NPB{153}{79}{61}, \\ 
\PLB{82}{79}{60};\\
E. Cremmer, J. Scherk and J.H. Schwarz, \PLB{84}{79}{83};\\
P. Fayet, \PLB{159}{85}{121}, \NPB{263}{86}{649}.

\bibitem{r} R. Rohm, \NPB{237}{84}{553}.

\bibitem{KP} C. Kounnas and M. Porrati, \NPB{310}{88}{355};\\ 
S. Ferrara,
C. Kounnas, M. Porrati and F. Zwirner, \NPB{318}{89}{75};\\ 
C. Kounnas and B. Rostand, \NPB{341}{90}{641};\\
I. Antoniadis and C. Kounnas, \PLB{261}{91}{369};\\
E. Kiritsis and C. Kounnas, \NPB{503}{97}{117}. 

\bibitem{DIN}  J.-P. Derendinger, L.E. Ib{\'a}{\~n}ez and H.P. Nilles,
  \PLB{155}{85}{65};\\
M. Dine, R. Rohm, N. Seiberg and E. Witten, \PLB{156}{85}{55}.

\bibitem{S} A. Sagnotti, in: Cargese '87, Non-Perturbative Quantum Field Theory,\\
eds. G. Mack et al. (Pergamon Press, Oxford, 1988) p. 521. 

\bibitem{W1} E. Witten, \NPB{443}{95}{85}, {\bf B471} (1996) 135.

\bibitem{PW} J. Polchinski and E. Witten, \NPB{460}{96}{525}.

\bibitem{HW} P. Horava and E. Witten, \NPB{460}{96}{506}, {\bf B475} (1996) 94.

\bibitem{AQ} I. Antoniadis and M. Quir\'os, \PLB{392}{97}{61}, \\
\NPB{505}{97}{109}, \PLB{416}{98}{327}.

\bibitem{DG} E. Dudas and C. Grojean, \NPB{507}{97}{553};\\ E. Dudas,
\PLB{416}{98}{309}.

\bibitem{Ba} C. Bachas, hep-th/9503030;\\
J.G. Russo and A.A. Tseytlin, \NPB{461}{96}{131}, hep-th/9804076.

\bibitem{PS} G. Pradisi and A. Sagnotti, \PLB{216}{89}{59}.

\bibitem{BS} M. Bianchi and A. Sagnotti, \PLB{247}{90}{517},\\  \NPB{361}{91}{519}.

\bibitem{PSS} G. Pradisi, A. Sagnotti and Ya.S. Stanev, 
\PLB{354}{95}{279}, \\{\bf B356} (1995) 230, {\bf B381} (1996) 97;\\
J. Fuchs and C.Schweigert, \PLB{414}{97}{251}.

\bibitem{BPS} M. Bianchi, G. Pradisi and A. Sagnotti, \NPB{376}{92}{365}.

\bibitem{BD} J. Blum and K. Dienes, \PLB{414}{97}{260}, \NPB{516}{98}{83}.

\bibitem{N} K.S. Narain, \PLB{169}{86}{41};\\ K.S. Narain, M.H. Sarmadi
and E. Witten, \NPB{279}{87}{369}.

\bibitem{GP} E. Gimon and J. Polchinski, hep-th/9601038.

\bibitem{six}{A. Dabholkar and J. Park, \NPB{472}{96}{207}, {\bf B477} (1996) 701;\\
E. Gimon and C.V. Johnson, \NPB{477}{96}{715};\\
M. Berkooz, R. Leigh, J. Polchinski, J.H.
Schwarz, N. Seiberg and E. Witten, \\{\sl Nucl. Phys.} {\bf B475} (1996) 115;\\
J. Blum and A. Zaffaroni, {\sl Phys. Lett.} {\bf B387} (1996) 71;\\ J. Blum, {\sl
Nucl. Phys.} {\bf B486} (1997) 34;\\ A. Sen, {\sl Nucl. Phys.} {\bf B475} (1996)
562,  \\{\sl Phys. Rev.} {\bf D55} (1997) 7345, hep-th/9709159;\\  
J. Blum and K. Intriligator, \NPB{506}{97}{199}, {\bf B506} (1997) 223;\\
K. Dasgupta and S. Mukhi, {\sl Phys. Lett.} {\bf B385} (1996) 125;\\
C. Angelantonj, M. Bianchi, G. Pradisi, A. Sagnotti and Ya.S. Stanev, \\{\sl
Phys. Lett.} {\bf B387} (1996) 743;\\Z. Kakushadze, G. Shiu and S.-H. Tye,
hep-th/9803141.}

\bibitem{AGW} L. Alvarez-Gaum\'e and E. Witten, \NPB{234}{84}{269}.

\bibitem{ABPSS} C. Angelantonj, M. Bianchi, G. Pradisi, A. Sagnotti
  and Ya. S. Stanev, \\
\PLB{385}{96}{96};\\
Z. Kakushadze, \NPB{512}{98}{221};\\
Z. Kakushadze and G. Shiu, \PRD{56}{97}{3686}, \\\NPB{520}{98}{75};\\
M. Bianchi, G. Pradisi and A. Sagnotti, 1991, as reported in:\\
A. Sagnotti, hep-th/9302099;\\
G. Zwart, hep-th/9708040;\\
G. Aldazabal, A. Font, L.E. Ib{\'a}{\~n}ez and G. Violero, hep-th/9804026;\\
Z. Kakushadze, G. Shiu and S.-H.H. Tye, hep-th/9804092.

\end{thebibliography}
\end{document}